\documentclass[reqno,11pt]{amsart}
\usepackage{defs}
\newcommand\figref{Figure~\ref}
\usepackage{apalike}
\newenvironment{spmatrix}[1]
 {\def\mysubscript{#1}\mathop\bgroup\begin{pmatrix}}
 {\end{pmatrix}\egroup_{\textstyle\mathstrut\mysubscript}}

\newcommand{\rate}[3]{\frac{\partial {#1}_{#3}}{\partial{#2}}}

\newcommand{\rates}[3]{\frac{\partial^2 {#1}_{#3}}{\partial {#2}^2}}
\newcommand{\rat}[2]{\frac{#1}{#2}}
\newcommand{\visp}[1]{{\bar{\mu}_{#1}}}
\newcommand{\dev}[2]{\frac{\partial \bar{#1}}{\partial \bar{#2}}}
\newcommand{\devu}[3]{\frac{\partial \bar{#1}_{#3}}{\partial \bar{#2}}}
\newcommand{\devus}[3]{\frac{\partial^2 \bar{#1}_{#3}}{\partial \bar{#2}^2}}
\calclayout

\makeatletter
\renewcommand{\author}[2][]{%
  \def\@tempa{#1}
  \ifx\@empty\authors
    \ifx\@tempa\@empty
      \gdef\shortauthors{#2}%
    \else
      \gdef\shortauthors{#1}%
    \fi
    \gdef\authors{\author{#2}}%
  \else
    \ifx\@tempa\@empty
      \g@addto@macro\shortauthors{\and#2}%
    \else
      \g@addto@macro\shortauthors{\and#1}%
    \fi
    \g@addto@macro\authors{\and\author{#2}}%
  \fi
}
\renewcommand{\address}[2][]{\g@addto@macro\authors{\address{#1}{#2}}}
\def\@setauthors{%
  \begin{center}%
    \footnotesize
    \vspace{20pt}
    \let\and\@empty
    \def\author##1{\advance\@tempcnta\@ne}%
    \def\address##1##2{\advance\@tempcntb\@ne}%
    \@tempcnta=\z@  \@tempcntb=\z@
    \authors
    \ifnum\@tempcnta>\@ne \ifnum\@tempcntb=\@ne
        \oneaddress
      \else
        \sepaddresses
      \fi
    \else
      \oneaddress
    \fi
  \end{center}%
}
\def\oneaddress{%
  \begingroup
  \let\author\@iden \let\address\@gobbletwo
  \renewcommand{\andify}{%
    \nxandlist{\unskip, }{\unskip{} and~}{\unskip, and~}}%
  \uppercasenonmath\authors
  \andify\authors
  \authors
  \endgroup
  \begingroup \let\and\relax \let\author\@gobble
  \def\address##1##2{\unskip\\[10pt] \itshape##2}%
  \authors
  \endgroup
}
\def\sepaddresses{%
  \begingroup
    \baselineskip10\p@\relax
    \def\address##1##2{ ({\itshape##2}\/)}
    \def\author##1{\def\temp{##1}\leavevmode\uppercasenonmath\temp\temp}%
    \nxandlist
      {,\\[\baselineskip]}
      {\\[\baselineskip] \textsc{\lowercase{and}}\\[\baselineskip]}
      {,\\[\baselineskip]\textsc{\lowercase{and}}\\[\baselineskip]}
      \authors 
    \authors
  \endgroup
}
\def\maketitle{\par
  \@topnum\z@
  \@setcopyright
  \thispagestyle{firstpage}%
  \uppercasenonmath\shorttitle
  \ifx\@empty\shortauthors \let\shortauthors\shorttitle
  \else
    \newcommand{\@xuppercasenonmath}[1]{\toks@\@emptytoks
      \@xp\@skipmath\@xp\@empty##1$$%
      \edef##1{\@nx\protect\@nx\@upprep\the\toks@}}%
    \@xuppercasenonmath\shortauthors
    \def\@@and{AND}
    \renewcommand{\andify}{%
      \nxandlist{\unskip, }{\unskip{ }\@@and{ }}{\unskip, \@@and{ }}}%
    \andify\shortauthors
  \fi
  \@maketitle@hook
  \begingroup
  \@maketitle
  \endgroup
  \c@footnote\z@
  \@cleartopmattertags
}
\def\@maketitle{%
  \normalfont\normalsize
  \let\@makefntext\noindent
  \@adminfootnotes
  \ifx\@empty\addresses\else \@footnotetext{\@setotheraddresses}\fi
  \global\topskip68\p@\relax
  \@settitle
  \ifx\@empty\authors \else \@setauthors \fi
  \ifx\@empty\@dedicatory
  \else
    \baselineskip26\p@
    \vtop{\centering{\footnotesize\itshape\@dedicatory\@@par}%
      \global\dimen@i\prevdepth}\prevdepth\dimen@i
  \fi
  \toks@\@xp{\shortauthors}\@temptokena\@xp{\shorttitle}%
  \edef\@tempa{\@nx\markboth{\the\toks@}{\the\@temptokena}}\@tempa
  \@setabstract
  \normalsize
  \if@titlepage
    \newpage
  \else
    \dimen@34\p@ \advance\dimen@-\baselineskip
    \vskip\dimen@\relax
  \fi
} 
\renewcommand{\thanks}[1]{%
  \ifx\@empty\thankses
    \gdef\thankses{\thanks{#1}}%
  \else
    \g@addto@macro\thankses{\endgraf\thanks{#1}}%
  \fi}
\def\@setthanks{\def\thanks##1{\noindent##1\@addpunct.}\thankses}

\renewcommand{\curraddr}[2][]{%
  \ifx\@empty\addresses
    \gdef\addresses{\curraddr{#1}{#2}}%
  \else
    \g@addto@macro\addresses{\endgraf\curraddr{#1}{#2}}%
  \fi}
\renewcommand{\email}[2][]{%
  \ifx\@empty\addresses
    \gdef\addresses{\email{#1}{#2}}%
  \else
    \g@addto@macro\addresses{\endgraf\email{#1}{#2}}%
  \fi}
\renewcommand{\urladdr}[2][]{%
  \ifx\@empty\addresses
    \gdef\addresses{\urladdr{#1}{#2}}%
  \else
    \g@addto@macro\addresses{\endgraf\urladdr{#1}{#2}}%
  \fi}
\def\@setotheraddresses{%
  \def\curraddr##1##2{\noindent
    \emph{Current address\@ifnotempty{##1}{ of ##1}}:\space
      ##2\@addpunct.}%
  \def\email##1##2{\noindent
    \emph{E-mail\@ifnotempty{##1}{ of ##1}}:\space
      \texttt{##2}}%
  \def\urladdr##1##2{\noindent
    \emph{WWW address\@ifnotempty{##1}{ of ##1}}:\space
      \texttt{##2}}%
  \addresses
}
\let\enddoc@text\relax
\makeatother
\begin{document}
\title[Two phase blood flow ] {MHD PULSATILE TWO-PHASE BLOOD FLOW THROUGH A STENOSED ARTERY WITH HEAT AND MASS 
TRANSFER}
\author{Bhavya Tripathi$^{{1}{*}}$, Bhupendra Kumar Sharma$^{1}$ \\}
\email{${^*}$(\textit{Corresponding author}){bhaviiitr2013@gmail.com}}
 \address{$^{1}$ Department of Mathematics, Birla Institute of Technology and Science, Pilani, Pilani Campus, Rajasthan-333031, India
}

\email{bksharma@pilani.bits-pilani.ac.in}
\author{Madhu Sharma$^{2}$\\}
\address{$^{2}$ Department of Bioscience, CASH, Mody University of Science and Technology, Lakshmangarh, Rajasthan-332311, India}
\begin{abstract}
In this paper, effects of heat and mass transfer on two-phase pulsatile blood flow through a narrowed stenosed artery with radiation and the chemical reaction has been investigated. A vertical artery is assumed in which magnetic field is applied along the radial direction of the artery.   The characteristics of blood in narrow arteries are analyzed by considering blood as Newtonian fluid in both core as well as in plasma regions. Exact solutions have been found for velocity, energy and concentration equations of the blood flow.  To understand the behavior of blood flow, graphs of velocity profile, wall shear stress,  flow rate, flow impedance and concentration profile have been portrayed for different values of the magnetic and radiation parameter.    In order to validate our result, a comparative study has been presented between the single-phase and two-phase model of the blood flow and it is observed that the two-phase model fits more accurately with the experimental data  than the single phase model. For pulsatile flow, the phase difference between the pressure gradient and flow rate has been displayed with magnetic field parameter and height of the stenosis.  Contour plots have been plotted for $10\%$, $20\%$ and $30\%$ case of an arterial blockage.
\subsection*{Keywords:}  Two-phase flow, stenosis, magnetohydrodynamic (MHD), chemical reaction, radiation, pulsatile flow.
\end{abstract}

\maketitle
\section{Introduction}
 The study of blood flow through arteries is of considerable importance in many cardiovascular diseases.   A major characteristic of the blood depends upon the hematocrit level as it is the percentage of whole blood occupies by the red blood cells(RBC)\cite{medvedev2011two}. Hematocrit level of the blood flow in the artery depends upon the diameter of the artery, and this relationship has important implications for physiological phenomena related to blood flow. Fahraeus effect explains this relationship which states that as the value of arterial diameter decreases, the hematocrit level present in the artery also decreases. In the arteries having the diameter less than 500 $\mu m$, erythrocytes move towards the center of the artery and thus forming the cell depleted plasma layer near the wall \cite{barbee1971fahraeus, faahraeus1931viscosity} due to the Fahraeus effect. In smaller arteries, the existence of cell-free plasma layer near the arterial wall and the higher concentration of red blood cell near the center, blood flow is considered as two-phase fluid flow \cite{verma1analytical}.
 Cokelet and Goldsmith \cite{cokelet1991decreased} through \textit{In vitro} analysis found that the two-phase flow of a suspension may lead to a decrease in hydrodynamic resistance for a tube of 172-$\mu$m diameter. Saran and Popel \cite{sharan2001two} presented a two-phase model of blood flow in narrow arteries assuming different viscosity of core region and plasma regions. Through their work, it is noticed that the effective viscosity of the plasma region increases as the level of the hematocrit increases due to the dissipation of energy.
 
  Atherosclerosis or stenosis is the important cause of death in various countries.  There is a significant proof that vascular fluid mechanics plays a key role in the progression and development of arterial stenosis, which is one of the most common diseases in mammals \cite{ku1997blood}. 
 Cardiovascular system transports nutrients and waste product from one body part to other. To supply proper oxygen-rich blood to all the tissues through arteries an adequate blood circulation is necessary \cite{alexopoulos2014visceral}. Many researchers have studied various aspects of blood flow in normal diseases arteries and blood flow in arteries having a single stenosis. To analyze the two-phase blood flow behavior under these hemodynamical disturbances, Ponalagusamy \cite{ponalagusamy2016two} stabilized the two-fluid model for blood flow through a tapered stenotic artery considering core region as couple stress fluid and a peripheral region of plasma as a Newtonian fluid. The author reported that wall shear stress is high in the case of converging tapered stenosis and it is low in the case of non-tapered diverging stenosis.  Further, for the case of symmetric and axisymmetric stenosis, Sankar \cite{sankar2011two} proposed a mathematical model for two-phase blood flow and investigated that presence of cell-depleted peripheral layer near the wall helps in the functioning of the diseased arterial system. Many authors explored the two-phase blood flow modeling of the stenosed artery considering the extra effect of the magnetic field.

 Magnetohydrodynamics (MHD) is to study the motion of highly conducting fluid under the influence of the magnetic field. The magnetic field with Newtonian fluid and non-Newtonian fluids has wide applications in chemical engineering, bio-fluid mechanics and various industries. If a magnetic field is applied to a moving electrically conducting liquid, it induces electric and magnetic fields. Haik \textit{et al.} \cite{haik2001apparent} reported a $30\%$ decrease in blood flow rate when subjected to a high magnetic field of $10 T$ while Yadav \textit{et al.} \cite{yadav2008experimental} showed a similar reduction in blood flow rate but at a much smaller magnetic field of $0.002 T$. The interaction of these fields produces a body force known as the Lorentz force which has a tendency to oppose the movement of the liquid \cite{shit2014mathematical,mekheimer2008effect,mekheimer2008influence}.  In the artery to recognize the existence of the stenosis,
a non-invasive technique based on MRI devices is used \cite{cavalcanti1995hemodynamics}. The MRI device uses the strong magnetic field to work and when performs over the particular area of our body it affects the velocity field of that area. Many researchers have been shown the effects of magnetization on the arterial vessel by considering the two-phase fluid model of the blood flow. Ponalagusamy and Selvi \cite{ponalagusamy2015influence} examined the effect of magnetic field on the two-phase model of oscillatory blood flow assuming both core and plasma regions as a Newtonian fluid in the presence of an arterial stenosis. The authors reported in their study that as the value of the magnetic field increases, the flow resistance of the blood flow in the stenosed artery also increases. Further, Ponalagusamy and Priyadharshini \cite{ponalagusamy2016numerical} discussed the effect of magnetic field on two-phase blood flow through a tapered stenosed artery, assuming micropolar fluid in the core region and Newtonian fluid in the plasma region. A mathematical model presented by Mirza \textbf{et al.} \cite{mirza2016transient} analyzed the effect of magnetic field on a transient laminar electromagneto-hydrodynamic two-phase blood flow using a continuum approach. Solving the model analytically they displayed the effect of the magnetic field for both blood velocity and particles velocity separately and concluded that as the effects of the magnetic field increases, both blood and particles velocities decrease for electromagneto- hydrodynamic two-phase blood flow. 

 The radiation effect in blood is of significant interest to clinicians in the therapeutic procedure of hyperthermia, which has a well-recognized effect in oncology. In pathological situations, thermal radiation therapy is one of the treatments employed by the medical practitioners \cite{sapareto1984thermal, damianou1993focal}.  Its effect is achieved by overheating the cancerous tissues \cite{szasz2007hyperthermia} by means of electromagnetic radiation. The procedure involves transmitting heat below the skin surface into tissues and muscles. Deep heat speeds up healing by increasing blood flow to the injury. In the past few decades, the study of the aggregated effects of heat and mass transfer on bio-fluids has got quite fascinating to the researchers both from the theoretical and observational or clinical perspective. Heat and mass transfer of blood flow considering its pulsatile hydromagnetic rheological nature under the presence of viscous dissipation and Joule heating discussed by Sharma \textit{et al.} \cite{sharma2013heat, sharma2015mathematical}. Sinha and Shit \cite{sinha2015electromagnetohydrodynamic} have investigated the combined effects of thermal radiation and MHD heat transfer blood flow through a capillary. Recently, Tripathi and Sharma \cite{tripathi2018influence} discussed the effect of heat and mass transfer on the two-phase model of the blood flow through a horizontal stenosed artery. The blood flow is considered as Newtonian fluid in both the core and the plasma region.

 There are many important technological problems that concern the flow of chemically-reacting fluid mixtures. Many biological fluid systems are examples of such mixtures. For example, blood is a complex mixture of plasma, proteins, cells, and a variety of other chemicals which is modeled usually in a homogenized sense as a single constituent fluid. Blood is maintained in a delicate balance by a variety of chemical reactions, some that aid its coagulation and others its dissolution.  Coulson and Hernandez \cite{hernandez1983alligator} carried out an experimental study that depicts the metabolic activities in the presence of chemical reactions. They made an important observation due to the injection of certain drugs, hormones and metabolites, and there occurs a considerable increase in the plasma concentration of reactants.  Xu \textit{et al.} \cite{xu2009study} developed a theoretical model to study the impact of blood flow on the growth of thrombi, by considering the interaction between different constituents of blood and chemical reaction. Recently, Sharma and Gaur \cite{sharma2017effect}  studied the effect of variable viscosity on chemically reacting magneto-blood flow with heat and mass transfer.

 Above mentioned studies, either were focused on analyzing momentum and heat transfer phenomenon for two-phase models of blood flow or mass transfer for a single phase with the horizontal artery. To the best knowledge of authors, combined effects of heat and mass transfer on MHD two-phase blood flow through the stenosed artery with radiation and chemical reaction have not been examined earlier. Therefore, the present article analyzes the effects of radiation, chemical reaction and the external applied the magnetic field on the two-phase of blood flow considering mild stenosis in the artery. The exact solutions for the velocity, temperature and concentration have been found out for both core and plasma regions. To get physical insight into the problem, velocity profile, flow resistance, total flow rate, wall shear stress, temperature and concentration profiles have been examined taking different values of magnetic field, the ratio of the viscosity parameter and radiation parameter. A comparative study has been done with experimental data to show the effectiveness of the two-phase model of blood flow and it is observed that the two-phase model fits more appropriately with the experimental data as compared to single phase model.

\section{The Mathematical Formulation}
\begin{figure}[h]
    \begin{center}    
       \includegraphics[width=6.6cm,height=40cm,keepaspectratio,angle =90]{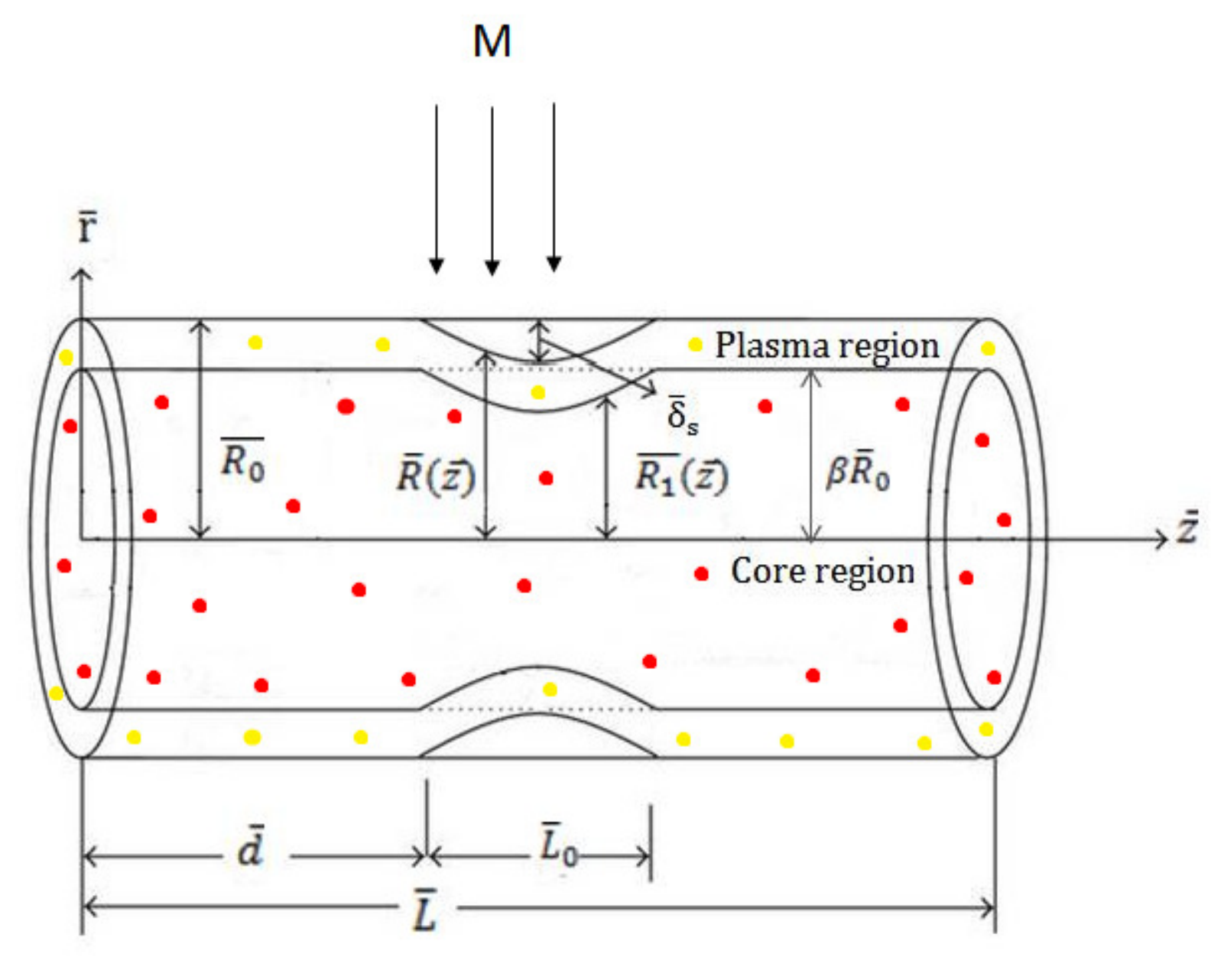} \hspace{-0.3in}       
\setlength{\belowcaptionskip}{-90ex}
       \hspace{-0.90in}\parbox{10in}{\caption{Geometry of the vertical stenosed artery of length $\bar{L}$ } \label{fig:artery}}
   \end{center}
\end{figure}

Consider the continuum model of unsteady, incompressible, pulsatile two-phase blood flow through a vertically stenosed coronary artery of length $L$ in the presence of applied magnetic field $M$ as shown in \figref{fig:artery}.  In the cylindrical artery of radius $r$, the two-phase model of blood flow consists of a core region of radius $r_c$ which contains erythrocytes (a suspension of the uniform hematocrit of viscosity $\bar{\mu_c}$ ) and a plasma region of radius $r_p$  having  the cell-depleted plasma layer with viscosity $\bar{\mu_p}$. The viscosities of the fluid for core and plasma regions  are given by
$$\bar{\mu}(\bar{r})=\left\{
\begin{array}{l}
\bar{\mu_c}\quad \text{for} \quad 0\leq {r} \leq  {r_c}(\bar{z}),\\
\bar{\mu_p} \quad \text{for} \quad  {r_c}(\bar{z}) \leq {r_p}(\bar{z}).
\end{array}
\right.$$
Note that the artery is assumed to be of cylindrical shape with $(\bar{u_c}, \bar{v_c}, \bar{w_c})$ as the velocity components for core region and $(\bar{u_p}, \bar{v_p}, \bar{w_p})$ as the velocity components for plasma region along $(\bar{r}, \bar{\theta}, \bar{z})$ direction in the cylindrical coordinates. Shear stresses are considered high enough so that  the fluid can be treated as Newtonian in both the regions \cite{sharan2001two}. Let $\bar{T_w}$ and $\bar{C_w}$ be the temperature  and concentration of the outer wall of the artery, respectively. The temperature  $\bar{T_w}$ is assumed high enough at the wall to induce radiative heat transfer.  

	Geometry of the stenosis in plasma region, which is assumed symmetric about the radial direction, is defined as \cite{mekheimer2008micropolar},
\begin{align}\label{stenosis_p}
\frac{\bar{R}(\bar{z})}{\bar{R_0}}=
\begin{cases}
1-{\frac{{\bar{\delta}_s}{n^{\frac{n}{n-1}}}}{{\bar{R}_0}{\bar{L}_0}^n\left(n-1\right)}}\left({\bar{L}_0}^{n-1}(\bar{z}-\bar{d})-(\bar{z}-\bar{d})^n\right) \quad \text{for} \quad & \bar{d}\leq \bar{z} \leq \bar{d}+\bar{L}_0,\\
1, & \text{otherwise},
\end{cases}
\end{align}
and in core region geometry of the stenosis is defined as \cite{sankar2007two},
\begin{align}\label{stenosis_c}
\frac{\bar{R_1}(\bar{z})}{\bar{R_0}}=
\begin{cases}
\beta-{\frac{{\bar{\delta}_s}{n^{\frac{n}{n-1}}}}{{\bar{R}_0}{\bar{L}_0}^n\left(n-1\right)}}\left({\bar{L}_0}^{n-1}(\bar{z}-\bar{d})-(\bar{z}-\bar{d})^n\right) \quad \text{for} \quad & \bar{d}\leq \bar{z} \leq \bar{d}+\bar{L}_0,\\
\beta, & \text{otherwise},
\end{cases}
\end{align}
 where $\bar{L}_0$ is the length of the stenosis, $\beta$ is the ratio of the central core radius to the normal artery radius, $\bar{R_0}$ is the radius of the artery and $n$ determines the shape of the constriction profile \cite{nadeem2012influence} and $\bar{\delta_s}$ indicates the maximum height of the stenosis located at
\begin{align}
\bar{z}=\bar{d}+\frac{\bar{L_0}}{n^{\frac{1}{\left(n-1\right)}}}.
\end{align}
For the symmetric case, i.e., $n=2$, maximum height of the stenosis occurs at mid point of the stenotic region
$$\bar{z}=\bar{d}+\frac{\bar{L_0}}{2}.$$
 
 The fluid flow in both the regions is in the axial direction and the applied magnetic field is acting perpendicular to the flow direction. Under these assumptions, the equations for momentum, energy and concentration for the core region are given by \label{core model}
\begin{align}\label{mom_c}
{\bar{\rho_c}}\devu{u}{t}{c}=-\dev{p}{z}+\visp{c}\left(\devus{u}{r}{c}+{\frac{1}{\bar{r}}}\devu{u}{r}{c}\right)-{\bar{\sigma}}{\bar{B_0}}^2{\bar{u_c}}+{\bar{g}}{\bar{\rho_c}}{\bar{\beta}}\left(\bar{T_c}-\bar{T_{0}}\right)+{\bar{g}}{\bar{\rho_c}}{\bar{\gamma}}\left(\bar{C_c}-\bar{C_{0}}\right), 
\end{align}
\begin{align}\label{tem_c}
\bar{\rho_c}\bar{c_c}\devu{T}{t}{c}=\bar{K_c}\left(\devus{T}{r}{c}+\frac{1}{\bar{r}}\devu{T}{r}{c}\right)-\devu{q}{r}{c},
\end{align}
\begin{align}\label{con_c}
\devu{C}{t}{c}=\bar{D_c}\left(\devus{C}{r}{c}+\frac{1}{\bar{r}}\devu{C}{r}{c}\right)-\bar{E'_c}\left({\bar{C_c}-{\bar{C_{{0}}}}}\right),
\end{align}
where $\bar{B_0}$ is the magnetic field intensity, $\bar{\sigma}$ is the electrical conductivity, $\dev{\bar{p}}{\bar{z}}$ represents the pressure gradient, $\bar{D_c}$ is the coefficient of mass diffusivity, $\bar{\rho_c}$ is the density, $\bar{c_c}$ is the specific heat, $\bar{E'_c}$ is the chemical reaction parameter, $\bar{K_c}$ is the thermal conductivity,  $\bar{u_c}$ is the velocity of the fluid in radial direction, $\bar{T_c}$ is the temperature and $\bar{C_c}$ is the concentration of fluid(blood) in the core region.

Similarly, the governing equations of momentum, energy and concentration for plasma region are given by\label{plasma model}
\begin{align}\label{mom_p}
{\bar{\rho_p}}\devu{u}{t}{p}=-\dev{p}{z}+\visp{p}\left(\devus{u}{r}{p}+{\frac{1}{\bar{r}}}\devu{u}{r}{p}\right)-{\bar{\sigma}}{\bar{B_0}}^2{\bar{u_p}}+{\bar{g}}{\bar{\rho_p}}{\bar{\beta}}\left(\bar{T_p}-\bar{T_{0}}\right)+{\bar{g}}{\bar{\rho_p}}{\bar{\gamma}}\left(\bar{C_p}-\bar{C_{0}}\right), 
\end{align}
\begin{align}\label{tem_p}
{\bar{\rho_p}}\bar{c_p}\devu{T}{t}{p}=\bar{K_p}\left(\devus{T}{r}{p}+\frac{1}{\bar{r}}\devu{T}{r}{p}\right)-\devu{q}{r}{p},
\end{align}
\begin{align}\label{con_p}
\devu{C}{t}{p}=\bar{D_p}\left(\devus{C}{r}{p}+\frac{1}{\bar{r}}\devu{C}{r}{p}\right)-\bar{E'_p}\left({\bar{C_p}-
{\bar{C_{{0}}}}}\right),
\end{align}
where $\bar{D_p}$ is the coefficient of mass diffusivity, $\bar{\rho_p}$ is the density, $\bar{c_p}$ is the specific heat, $\bar{E'_p}$ is the chemical reaction parameter and $\bar{K_p}$ is the thermal conductivity of  plasma in the peripheral region. 
\begin{remark}
Terms $\devu{q}{r}{c}$ in Eq. \eqref{tem_c} and  $\devu{q}{r}{p}$  in Eq. \eqref{tem_p} are due to the radiation effect of heat transfer, where $\bar{q_c}$ and $\bar{q_p}$ represent the radiative heat flux in core and plasma regions respectively.
\end{remark}
The conservation equation of the radiative heat transfer for all wavelength in a unit volume  is described as\cite{cogley1968differential}
\begin{align}\label{rad}
\nabla.{\bar{q_c}}=\int_{0}^{\infty} {K_{\lambda}(\bar{T})(4{e_{\lambda{h}}}(\bar{T})-{G_{\lambda}})} d{\lambda},
\end{align}
where ${e_{\lambda{h}}}$ is the Plank's function, $G_{\lambda}$ is the incident radiation, i.e.,  
\begin{align*}
G_{\lambda}=\frac{1}{\pi}\int_{4{\pi}} {e_{\lambda}{(\Omega)}} d{\Omega},
\end{align*}
and the injection parameter $\lambda$  is given by\cite{sharma2007radiation}
\begin{align*}
\lambda=\frac{{u_0}{R_0}}{\mu}.
\end{align*}
Now, according to the Kirchhoff's law, the conserved radiative heat transfer Eq. \eqref{rad} for an incident radiation $G_{\lambda}=4{e_{\lambda{h}}(T_0)}$ specializes to
\begin{align}\label{Rad2}
\nabla.{\bar{q_c}}=4\int_{0}^{\infty} {K_{\lambda}(\bar{T})({e_{\lambda{h}}}(\bar{T})-{e_{\lambda{h}}}\bar{(T_0)}}) d{\lambda}.
\end{align}Subsequently, using Taylor series expansion of ${K_{\lambda}}(\bar{T})$ and $e_{\lambda{h}}({T_0})$, Eq. \eqref{Rad2} can be written as
\begin{align}\label{Rad3}
\nabla.{\bar{q_c}}=4({\bar{T}-{\bar{T_0}}})\int_{0}^{\infty}{K_{\lambda_0}{\left(\frac{{\partial}{{e_{\lambda{h}}}}}{\partial{T}}\right)}_0} d{\lambda},
\end{align}
where $K_{\lambda_0}=K_{\lambda}\bar{(T_0)}$.
Further, assuming ${{\alpha_c}^2}=\int_{0}^{\infty}{K_{\lambda_0}{\left(\frac{{\partial}{{e_{\lambda{h}}}}}{\partial{T}}\right)}_0} d{\lambda}$,  Eq. \eqref{Rad3} reduces to
\begin{align*}
\nabla.{\bar{q_c}}=4({\bar{T}-{\bar{T_0}}}){{\alpha_c}^2}.
\end{align*}
Therefore, the radiative heat fluxes for core and plasma regions respectively can be expressed  as
\begin{align}
\devu{q}{r}{c}=4{\bar{\alpha_c}}^2\left(\bar{T_c}-\bar{T_{{0}}}\right), \quad \devu{q}{r}{p}=4{\bar{\alpha_p}}^2\left(\bar{T_p}-\bar{T_{{0}}}\right),
\end{align}
where $\bar{q_c}$ and $\bar{q_p}$ represent the radiative heat transfer coefficients, and $\bar{\alpha_c}$ and
$\bar{\alpha_p}$ are the mean radiation absorption coefficients for core and plasma regions respectively. 
\begin{remark}
Note that the mean radiation absorption coefficients $bar{\alpha_c}$ and $\bar{\alpha_p}$ , in general, are considered less than unity ($\bar{\alpha}\ll 1$) in view of the fact that fluids like plasma and blood in the physiological conditions are optically thin with low density \cite{ogulu2005simulation}.
\end{remark} 
To solve the momentum, energy and concentration equations of the two-phase model of the blood flow no-slip boundary conditions are considered on the arterial wall. It is assumed that the functions of velocity, temperature and concentration are continuous at the interface of core and plasma regions so their values of both core and plasma regions are equal at that point and due to symmetry their gradient vanishes along with the axis. It is considered that at the interface of core and plasma regions heat and mass transfer effects are the same for both the regions\cite{sharan2001two}. The appropriate boundary conditions for the model under consideration are as follows:
\begin{equation} \label{eq:bnd}
\left\{
\begin{array}{l}
\bar{u_p}=0,\quad \bar{T_p}=\bar{T_w}, \quad \bar{C_p}=\bar{C_w}  \quad  at  \quad \bar{r}=\bar{R}(\bar{z})\\
{\bar{u_c}}={\bar{u_p}}, \quad \bar{T_p}=\bar{T_c}, \quad \bar{C_p}=\bar{C_c}  \quad  at  \quad \bar{r}=\bar{R_1}(\bar{z})\\
\devu{u}{r}{c}=0, \quad \devu{T}{r}{c}=0,   \quad \devu{C}{r}{c}=0  \quad  at  \quad \bar{r}=0\\
{\bar{\tau_c}}=\bar{\tau_p}, \quad \devu{T}{r}{c}=\devu{T}{r}{p}, \quad \devu{C}{r}{c}=\devu{C}{r}{p} \quad  at  \quad \bar{r}=\bar{R_1}(\bar{z}).
\end{array}\right.
\end{equation}
Now, introducing the following dimensionless parameters:
\begin{align*}
{{u_c}}\Let\frac{\bar{u_c}}{\bar{u_0}}, \quad {r}\Let\frac{\bar{r}}{\bar{R_0}},  \quad z\Let\frac{\bar{z}}{\bar{R_0}}, \quad t\Let{\bar{\omega}}{\bar{t}},\quad R(z) \Let\frac{\bar{R}(\bar{z})}{\bar{R_0}}, \quad {R_1}(z)\Let \frac{\bar{R}_1(\bar{z})}{\bar{R_0}}, 
\end{align*}
\begin{align*}
p\Let\frac{{\bar{R_0}}{\bar{p}}}{\bar{u_0}{\bar{\mu_p}} }, \quad {Re}\Let\frac{\bar{\rho_p}{{\bar{R_0}}^2}\omega}{\bar{\mu_p}},\quad {\theta_c}\Let\frac{(\bar{T_c}-\bar{T_{{0}}})}{\bar{T_w}-\bar{T_{{0}}}},  \quad{\sigma_c}\Let\frac{(\bar{C_c}-\bar{C_{{0}}})}{\bar{C_w}-\bar{C_{{0}}}}, \quad {{u_p}}\Let\frac{\bar{u_p}}{{\bar{u_0}}},
\end{align*}
\begin{align*}
\delta\Let\frac{\bar{\delta_s}}{\bar{R_0}}\quad{\theta_p}\Let\frac{(\bar{T_p}-\bar{T_{{0}}})}{\bar{T_w}-\bar{T_{{0}}}},\quad  {N^2}\Let\frac{4{\bar{R_0}}^2{\bar{\alpha_p}}^2}{\bar{K_p}}, \quad {M^2}\Let\frac{\bar{\sigma} {\bar{B_0}}^2{\bar{R_0}}^2}{\bar{\mu_p}}, \quad {\sigma_p}\Let\frac{(\bar{C_p}-\bar{C_{{0}}})}{\bar{C_w}-\bar{C_{{0}}}},
\end{align*}
\begin{align*}
{P_e}\Let\frac{\bar{\rho_p}{\bar{c_p}}{\bar{R_0}}^2{\bar{\omega}}}{\bar{K_p}}, \quad S_c\Let\frac{\bar{\mu_p}}{\bar{D_p}{\bar{\rho_p}}}, \quad {G_r}\Let\frac{\bar{g}{\bar{\rho_p}}{\bar{\beta}}{\bar{R_0}}^2\left({\bar{T_w}-\bar{T_{{0}}}}\right)}{\bar{u_0}{\bar{\mu_p}}}, \quad {D_0}\Let\frac{\bar{D_p}}{\bar{D_c}},
\end{align*}
\begin{align*}
{\tau_c}\Let\frac{\bar{\tau_c}{\bar{R_0}}^2}{\bar{u_0}{\bar{\mu_p}}}, \quad {\tau_p}\Let\frac{\bar{\tau_p}{\bar{R_0}}^2}{\bar{u_0}{\bar{\mu_p}}}, \quad {\rho_0}\Let\frac{\bar{\rho_p}}{\bar{\rho_c}}, \quad {\mu_0}\Let\frac{\bar{\mu_p}}{\bar{\mu_c}},  \quad \bar{E'_p}\Let\frac{E{\bar{\mu_p}}}{\bar{\rho_p}{\bar{R_0}}^2}, \quad {E_0}\Let\frac{\bar{E_p}}{\bar{E_c}}.
\end{align*}
The equations of momentum, energy and concentration Eqs. \eqref{mom_c}-\eqref{con_c} of the core region is represented in terms of these non-dimensional parameters as
\begin{align}\label{momntumC}
\left(\frac{{Re}}{\rho_0}\right)\rate{u}{t}{c}=-\frac{\partial{p}}{\partial{z}}+\frac{1}{\mu_0}\left(\rates{u}{r}{c}+\rat{1}{r}{\rate{u}{r}{c}}\right)-{M^2}{u_c}+\left(\rat{G_r}{\rho_0}\right){\theta_c}+\left(\rat{G_m}{\rho_0}\right){\sigma_c},
\end{align}
\begin{align}\label{temperatureC}
\frac{P_e{K_0}}{{\rho_0}{s_0}}\left({\rate{\theta}{t}{c}}\right)=\left(\rates{\theta}{r}{c}+\frac{1}{r}\rate{\theta}{r}{c}\right)-\frac{K_0}{\alpha_0}{N^2}{\theta_c},
\end{align}
\begin{align}\label{concentrationC}
{Re}\left(\rate{\sigma}{t}{c}\right)=\frac{1}{D_0}\left(\frac{1}{S_c}\right)\left(\rates{\sigma}{r}{c}+{\frac{1}{r}}\rate{\sigma}{r}{c}\right)-\frac{E}{E_0}{\sigma_c},
\end{align}
and, equations of momentum, energy and concentration Eqs. \eqref{mom_p}-\eqref{con_p} of the plasma region is represented in terms of these non-dimensional parameters as
\begin{align}\label{momentumP}
{Re}\rate{u}{t}{p}=-\frac{\partial{p}}{\partial{z}}+\left(\rates{u}{r}{p}+\rat{1}{r}{\rate{u}{r}{p}}\right)-{M^2}{u_p}+{G_r}{\theta_p}+{G_m}{\sigma_p},
\end{align}
\begin{align}\label{temperatureP}
{P_e}{\rate{\theta}{t}{p}}=\left(\rates{\theta}{r}{p}+\frac{1}{r}\rate{\theta}{r}{p}\right)-{N^2}{\theta_p},
\end{align}
\begin{align}\label{concentrationP}
{Re}\left(\rate{\sigma}{t}{p}\right)=\left(\frac{1}{S_c}\right)\left(\rates{\sigma}{r}{p}+\frac{1}{r}\rate{\sigma}{r}{p}\right)-E{\sigma_p},
\end{align}
where $\alpha_0$ is the ratio of the mean radiation absorption coefficient in the plasma
region to mean radiation absorption coefficient in the core region, $K_0$ is the ratio of thermal conductivity of plasma to core region, $s_0$ is the specific heat ratio of plasma to core region, $E_0$ is the ratio of chemical moles present in the plasma region to chemical moles in the core region  and $N$, $M$ and $E$ are factors of thermal radiation, applied magnetic field and chemical reaction for both core and plasma regions.

    The dimensionless form of the geometry of the stenosis in core and plasma regions are given by
\begin{align}
R({z})=
\begin{cases}
1-{\eta}\left(({z}-l)-({z}-l)^n\right) \quad \text{for} \quad &l\leq {z} \leq 1+l,\\
1  &\text{otherwise},
\end{cases}
\end{align}

\begin{align}
{R_1}({z})=
\begin{cases}
\beta-{\eta}\left(({z}-l)-({z}-l)^n\right) \quad \text{for} \quad & l\leq {z} \leq 1+l,\\
\beta & \text{otherwise},
\end{cases}
\end{align}
where $${\eta}=\frac{{{\delta}}{n^{\frac{n}{n-1}}}}{\left(n-1\right)},\quad l=\frac{\bar{d}}{\bar{L_0}}, \quad\delta=\frac{\bar{\delta_s}}{\bar{R_0}}.$$
and the corresponding boundary conditions Eq. \eqref{eq:bnd} in non-dimensional form for both core and plasma regions are given as
\begin{align}\label{BC}
\left\{
\begin{array}{l}
{{u_p}}=0,\quad\quad{{\theta}_p}=1,\quad\quad{{\sigma}_p}=1\quad at \quad \quad {r}={R}({z}),\\
{{u_p}}={{u_c}},\quad{{\theta}_p}={{\theta}_c},\quad\quad{{\sigma}_p}={{\sigma}_c}\quad at \quad\quad {r}={R_1}({z}),\\
{\tau_c}={\tau_p},\quad\rate{\theta}{r}{c}=\rate{\theta}{r}{p},\quad\rate{\sigma}{r}{c}=\rate{\sigma}{r}{p}\quad at \quad\quad {r}={R_1}({z}),\\
\rate{u}{r}{c}=0,\quad\rate{\theta}{r}{c}=0,\quad\quad\rate{\sigma}{r}{c}=0 \quad\quad at \quad\quad {r}=0.
\end{array}\right.
\end{align}

\section{Solution of the problem}
In view of the fact that the pumping action of the heart results in a pulsatile blood flow, pressure gradient can be represented as
\begin{align*}
-{\frac{\partial p}{\partial z}}= {P_0}{e^{i\bar{\omega} \bar{t}}},
\end{align*}
where $P_0$ represents the constant pressure. This assumption is limited to cases of harmonic oscillatory motion \cite{san2012dynamics}.

As Eqs. \eqref{momntumC}-\eqref{concentrationP} are linear, we are allowed to express velocity, temperature and concentration in the form of:
\begin{align}\label{withoutt}
\left\{
\begin{array}{l}
{u_c}(r,t)={u_{c_0}}(r){e^{i\omega t}}, \quad\quad {u_p}(r,t)={u_{p_0}}(r){e^{i\omega t}},\\
{\theta_c}(r,t)={\theta_{c_0}}(r){e^{i\omega t}},\quad\quad {\theta_p}(r,t)={\theta_{p_0}}(r){e^{i\omega t}},\\
{\sigma_c}(r,t)={{\sigma_{c_0}}(r){e^{i\omega t}}}, \quad\quad {\sigma_p}(r,t)={{\sigma_{p_0}}(r){e^{i\omega t}}}.
\end{array}\right.
\end{align}
 
Therefore, using Eq. \eqref{withoutt}, Eqs. \eqref{momntumC}-\eqref{concentrationC} can be reduced as
\begin{align}\label{mom_tc}
\left(\rates{u}{r}{c_0}+\rat{1}{r}{\rate{u}{r}{c_0}}\right)-\left({M^2}+\frac{{\mu_0}{Re}}{\rho_0}i\right){u_{c_0}}=-\left({P_0}+\frac{{G_r}{\theta_{c_0}}}{\rho_0}+\frac{{G_m}{\sigma_{c_0}}}{\rho_0}\right){\mu_0},
\end{align}
\begin{align}\label{tem_tc}
\rates{\theta}{r}{c_0}+\frac{1}{r}{\rate{\theta}{r}{c_0}}-\left(\frac{{K_0}{{N}^2}}{{\alpha_0}}+i{\frac{P_e}{\rho_0}\left(\frac{K_0}{s_0}\right)}\right){\theta_{c_0}}=0,
\end{align}
\begin{align}\label{con_tc}
\rates{\sigma}{r}{c_0}+\frac{1}{r}{\rate{\sigma}{r}{c_0}}-\left(i{Re}{D_0}{S_c}+\frac{E}{E_0}{D_0}{S_c}\right)\sigma_{c_0}=0.
\end{align}
Similarly, substituting Eq. \eqref{withoutt} into Eqs. \eqref{momentumP}-\eqref{concentrationP}, we get
\begin{align}\label{mom_tp}
\left(\rates{u}{r}{p_0}+\rat{1}{r}{\rate{u}{r}{p_0}}\right)-\left(M^2+{{Re}i}\right){u_{p_0}}=-\left({P_0}+\frac{{G_r}{\theta_{p_0}}}{\rho_0}+\frac{{G_m}{\sigma_{p_0}}}{\rho_0}\right),
\end{align}
\begin{align}\label{tem_tp}
\rates{\theta}{r}{p_0}+\frac{1}{r}{\rate{\theta}{r}{p_0}}-\left({{N}^2}+i{P_e}\right){\theta_{p_0}}=0,
\end{align}
\begin{align}\label{con_tp}
\rates{\sigma}{r}{p_0}+\frac{1}{r}{\rate{\sigma}{r}{p_0}}-\left(i{Re}{S_c}+{E}{S_c}\right)\sigma_{p_0}=0.
\end{align}
To find the solution of Eqs. \eqref{tem_tc} and \eqref{tem_tp} under the given boundary conditions Eq. \eqref{BC}, we rewrite the energy equations of core and plasma regions in standard  Bessel differential equations form as
\begin{align}\label{tem_tc1}
\rates{\theta}{r}{c_0}+\frac{1}{r}{\rate{\theta}{r}{c_0}}+{\beta_1}{\theta_{c_0}}=0,
\end{align}
\begin{align}\label{tem_tp1}
\rates{\theta}{r}{p_0}+\frac{1}{r}{\rate{\theta}{r}{p_0}}+{\beta_2}{\theta_{p_0}}=0,
\end{align}

 Hence, the solution of energy equations using Bessel functions for  core and plasma regions respectively, are
\begin{align}\label{thetac0}
{\theta_{c_0}(r)}=\left[{U_1}\left(\frac{\sqrt{\beta_2}{Y_1}(\sqrt{\beta_2}{R_1})}{U_3{Y_0(\sqrt{\beta_2}{R})}}-\frac{\sqrt{\beta_1}{U_2}{J_1}(\sqrt{\beta_1}{R_1})}{U_3}\right)+{U_2}\right]{{J_0}(\sqrt{\beta_1}(r))},
\end{align}
\begin{align}\label{thetapo}
{\theta_{p_0}(r)} = & \left[\left(\frac{\sqrt{\beta_2}{Y_1}(\sqrt{\beta_2}{R_1})}{U_3{Y_0(\sqrt{\beta_2}{R})}}-\frac{\sqrt{\beta_1}{U_2}{J_1}(\sqrt{\beta_1}{R_1})}{U_3}\right)\left({{J_0}(\sqrt{\beta_2}{r})}-\frac{J_0(\sqrt{\beta_2}{R}}{Y_0({\sqrt{\beta_2}}{R}}){Y_0({\sqrt{\beta_2}{r}})}\right)\right] \nonumber \\
& +\frac{Y_0({\sqrt{\beta_2}{r}})}{Y_0({\sqrt{\beta_2}{R}})}.
\end{align}
where $J_n(x)$ and $Y_n(x)$ are  respectively the Bessel function of first and second kind. Expressions for the constants ${U_1}$, ${U_2}$, ${U_3}$, $\beta_1$ and $\beta_2$ are given in Appendix A. 

 The expressions for temperature considering unsteady flow of core and plasma regions respectively, are as follows
\begin{align}
{\theta_c(r,t)}=\left[\left({U_1}\left(\frac{\sqrt{\beta_2}{Y_1}(\sqrt{\beta_2}{R_1})}{U_3{Y_0(\sqrt{\beta_2}{R})}}-\frac{\sqrt{\beta_1}{U_2}{J_1}(\sqrt{\beta_1}{R_1})}{U_3}\right)+{U_2}\right){{J_0}(\sqrt{\beta_1}(r))}\right]{e^{i\omega t}},
\end{align}
\begin{align}
{\theta_p(r,t)}=& \left[\left(\frac{\sqrt{\beta_2}{Y_1}(\sqrt{\beta_2}{R_1})}{U_3{Y_0(\sqrt{\beta_2}{R})}}-\frac{\sqrt{\beta_1}{U_2}{J_1}(\sqrt{\beta_1}{R_1})}{U_3}\right)\left({{J_0}(\sqrt{\beta_2}{r})}-\frac{J_0(\sqrt{\beta_2}{R}}{Y_0({\sqrt{\beta_2}}{R}}){Y_0({\sqrt{\beta_2}{r}})}\right)\right]{e^{i\omega t}} \nonumber \\
& +\frac{Y_0({\sqrt{\beta_2}{r}})}{Y_0({\sqrt{\beta_2}{R}})}{e^{i \omega t}}.
\end{align}
The concentration Eqs. \eqref{con_tc} and \eqref{con_tp} of core and plasma region can be  rewritten in the standard Bessel differential equation form as

\begin{align}
\rates{\sigma}{r}{c_0}+\frac{1}{r}{\rate{\sigma}{r}{c_0}}+{\gamma_1}\sigma_{c_0}=0,
\end{align}
\begin{align}
\rates{\sigma}{r}{p_0}+\frac{1}{r}{\rate{\sigma}{r}{p_0}}+{\gamma_2}\sigma_{p_0}=0,
\end{align}

The solutions of concentration equations using Bessel functions and satisfying the given boundary conditions Eq. \eqref{BC} for both core and plasma regions respectively, are 
\begin{align}\label{sigmac0}
{\sigma_{c_0}(r)}=\left[{U_4}\left(\frac{\sqrt{\gamma_2}{Y_1}(\sqrt{\gamma_2}{R_1})}{U_6{Y_0(\sqrt{\gamma_2}{R})}}-\frac{\sqrt{\gamma_1}{U_5}{J_1}(\sqrt{\gamma_1}{R_1})}{U_6}\right)+{U_5}\right]{{J_0}(\sqrt{\gamma_1}(r))},
\end{align}
and
\begin{align}
{\sigma_{p_0}(r)}= & \left[\left(\frac{\sqrt{\gamma_2}{Y_1}(\sqrt{\gamma_2}{R_1})}{U_6{Y_0(\sqrt{\gamma_2}{R})}}-\frac{\sqrt{\gamma_1}{U_5}{J_1}(\sqrt{\gamma_1}{R_1})}{U_6}\right)\left({{J_0}(\sqrt{\gamma_2}{r})}-\frac{J_0(\sqrt{\gamma_2}{R}}{Y_0({\sqrt{\gamma_2}}{R}}){Y_0({\sqrt{\gamma_2}{r}})}\right)\right]\\  \nonumber
& +\frac{Y_0({\sqrt{\gamma_2}{r}})}{Y_0({\sqrt{\gamma_2}{R}})},
\end{align}
Consequently, concentration for unsteady flow in core and plasma regions are as follows
\begin{align}
{\sigma_c(r,t)}=\left[{U_4}\left(\frac{\sqrt{\gamma_2}{Y_1}(\sqrt{\gamma_2}{R_1})}{U_6{Y_0(\sqrt{\gamma_2}{R})}}-\frac{\sqrt{\gamma_1}{U_5}{J_1}(\sqrt{\gamma_1}{R_1})}{U_6}\right)+{U_5}\right]{{J_0}(\sqrt{\gamma_1}(r))}{e^{i \omega t}},
\end{align}
\begin{align}
{\sigma_p(r,t)}= & \left[\left(\frac{\sqrt{\gamma_2}{Y_1}(\sqrt{\gamma_2}{R_1})}{U_6{Y_0(\sqrt{\gamma_2}{R})}}-\frac{\sqrt{\gamma_1}{U_5}{J_1}(\sqrt{\gamma_1}{R_1})}{U_6}\right)\left({{J_0}(\sqrt{\gamma_2}{r})}-\frac{J_0(\sqrt{\gamma_2}{R}}{Y_0({\sqrt{\gamma_2}}{R}}){Y_0({\sqrt{\gamma_2}{r}})}\right)\right]{e^{i \omega t}} \nonumber \\
& +\frac{Y_0({\sqrt{\gamma_2}{r}})}{Y_0({\sqrt{\gamma_2}{R}})}{e^{i\omega t}},
\end{align}
where expressions for the constants ${U_4}$, ${U_5}$, ${U_6}$, $\gamma_1$ and $\gamma_2$ are given in Appendix A.

 To find the solution for velocity profile in core region, we substitute the values of $\theta_{c_0}$ and $\sigma_{c_0}$ from Eq. \eqref{thetac0} and Eq. \eqref{sigmac0} into Eq. \eqref{mom_tc} and get the following non-homogeneous differential equations
\begin{align}\label{lambda1}
\left(\rates{u}{r}{c_0}+\rat{1}{r}{\rate{u}{r}{c_0}}\right)+{\lambda_1}{u_{c_0}}=-\left({P_0}+\frac{{G_r}{\theta_{c_0}}}{\rho_0}+\frac{{G_m}{\sigma_{c_0}}}{\rho_0}\right){\mu_0},
\end{align}
where expressions for $\lambda_1$ and $\lambda_2$ are given in Appendix A.

The solution of the Eq. \eqref{lambda1} is given as
\begin{align}\label{general solution1}
u_{c_0}= {C_1}{J_0{\sqrt{\lambda_1}r}}+{C_2}{Y_0{\sqrt{\lambda_1}r}}+ {A_1}{J_0{\sqrt{\lambda_1}r}}+{B_1}{Y_0{\sqrt{\lambda_1}r}},
\end{align}
where $C_1$, $C_2$ are the arbitrary constants and $A_1$, $A_2$ are defined as
\begin{align*}
A_1=-\frac{\pi r}{2}\int{Y_0{\sqrt{\lambda_1}r}\left({P_0}+\frac{{G_r}{\theta_{c_0}}}{\rho_0}+\frac{{G_m}{\sigma_{c_0}}}{\rho_0}\right){\mu_0}}\quad{dr},
\end{align*}
\begin{align*}
B_1=\frac{\pi r}{2}\int{J_0{\sqrt{\lambda_1}r}\left({P_0}+\frac{{G_r}{\theta_{c_0}}}{\rho_0}+\frac{{G_m}{\sigma_{c_0}}}{\rho_0}\right){\mu_0}}\quad{dr}.
\end{align*}
The Eq. \eqref{mom_tp} can be expressed as

\begin{align}\label{lambda2}
\left(\rates{u}{r}{p_0}+\rat{1}{r}{\rate{u}{r}{p_0}}\right)+\lambda_2{u_{p_0}}=-\left({P_0}+\frac{{G_r}{\theta_{p_0}}}{\rho_0}+\frac{{G_m}{\sigma_{p_0}}}{\rho_0}\right).
\end{align}
So, the solution of the given Eq. \eqref{lambda2} can be calculated as
\begin{align}\label{general solution2}
u_{p_0}={C_3}{J_0{\sqrt{\lambda_1}r}}+{C_4}{Y_0{\sqrt{\lambda_1}r}}+{A_2}{J_0{\sqrt{\lambda_1}r}}+{B_2}{Y_0{\sqrt{\lambda_1}r}},
\end{align}
where $C_3$, $C_4$ are arbitrary constants and $A_2$, $B_2$ are expressed as
\begin{align*}
{A_2}=-\frac{\pi r}{2}\int{Y_0{\sqrt{\lambda_2}r}\left({P_0}+\frac{{G_r}{\theta_{c_0}}}{\rho_0}+\frac{{G_m}{\sigma_{c_0}}}{\rho_0}\right)} \quad{dr},
\end{align*}
\begin{align*}
{B_2}=\frac{\pi r}{2}{\int{J_0{\sqrt{\lambda_2}r}\left({P_0}+\frac{{G_r}{\theta_{c_0}}}{\rho_0}+\frac{{G_m}{\sigma_{c_0}}}{\rho_0}\right)}}\quad{dr}.
\end{align*}

After applying the boundary conditions Eq. \eqref{BC} into the Eqs. \eqref{general solution1} and \eqref{general solution2}, we get ${C_2}=0$ and the linear system in terms of $C_1$, $C_3$, $C_4$ as
\begin{align}\label{Matrix}
\begin{spmatrix}{}
{J_0}(\sqrt{\lambda_1}{R_1}) & -{J_0}(\sqrt{\lambda_2}{R_1}) & -{Y_0}(\sqrt{\lambda_2}{R_1})\\      0 & {J_0}(\sqrt{\lambda_2}{R}) & {Y_0}(\sqrt{\lambda_2}{R})\\
 -{\sqrt{\lambda_1}}{J_1}(\sqrt{\lambda_1}{R_1}) & {\sqrt{\lambda_2}}{J_1}(\sqrt{\lambda_2}{R_1}) & {\sqrt{\lambda_2}}{Y_1}(\sqrt{\lambda_2}{R_1})
\end{spmatrix}
\begin{spmatrix}{}
    C_1 \\        C_3     \\C_4
\end{spmatrix}
=
\begin{spmatrix}{}
    D_1  \\        D_2      \\   D_3
\end{spmatrix}
\end{align}
where $D_1$, $D_2$ and $D_3$ are expressed in Appendix A.
Under the given set of boundary conditions the linear system Eq. \eqref{Matrix} admits a unique solution.
\section{Results and discussion}
For having adequate insight into the two-phase flow behavior of blood through a stenosed arterial segment, flow resistance, total flow rate, and wall shear stress have been estimated assuming pulsatile, Newtonian nature of the blood flow for both core and plasma regions. A computational study has been carried out to show the effects of cell-depleted plasma layer on blood flow with the variation of different quantities of interest. In all the figures, continuous lines show the respective profile for the core region while dotted lines display the same for the plasma region.
 Default values of the parameters used to graphically analyze the effectiveness of the model are given in Table.\ref{table 1}.
\begin{table}[h]
\centering
\begin{tabular}{ p{6cm} p{3cm} p{3cm}  }
 \hline
 \multicolumn{3}{c}{} \\
 Parameters                                          &Values(Unit free)    & Source\\
 \hline\\
Magnetic field Parameter ($M$)                       &  1.5-3             &  \cite{trivedi2004multi, swartz2009intracranial} \\
Schmidt Parameter ($S_c$)                            &   0.5-1.5          & \cite{zaman2016heat}\\
Radiation Parameter ($N$)                            & 2-5                & \cite{ogulu2005simulation}\\
Chemical reaction Parameter ($E$)                    & 0.5-2              & \cite{misra2016mhd}\\
Peclet Number ($P_e$)                                &  0.87              & \cite{sharan1997two}\\
Grashof Number ($G_r$)                               & 2-3                & \cite{misra2016mhd}\\
Modified Grashof Number ($G_m$)                      & 2-3                & \cite{misra2016mhd}\\
Ratio of Thermal Conductivity in core and plasma
regions ($K_0$)                                      & 0.4-0.8            & \cite{ponalagusamy2015influence} \\
Ratio of Specific heat in core and
plasma regions ($s_0$)                               & 1.0                & \cite{ponalagusamy2015influence} \\
Ratio of density in core and plasma
regions ($\rho_0$)                                   & 1.05                & \cite{medvedev2011two} \\
Ratio of mean radiation absobtion
coefficients in core and plasma region ($\alpha_0$)  & 1.0                & \cite{ponalagusamy2015influence} \\
Reynold Number($Re$)                                & 0.005              & \cite{fujiwara2009red, sharan1997two} \\
Ratio of viscosities in core and
plasma region ($\mu_0$)                              & 1.2                & \cite{medvedev2011two, sharan2001two} \\
Pressure gradient ($P_0$)                            &10                  &\cite{ponalagusamy2015influence}\\
 \hline
\end{tabular}
\setlength{\belowcaptionskip}{-1ex}
 \caption{{Values of the parameters}\label{table 1}}
 \end{table}

\begin{figure}[h]
  \centering
  \begin{minipage}[b]{0.49\textwidth}
    \includegraphics[width=\textwidth,height=5.5cm]{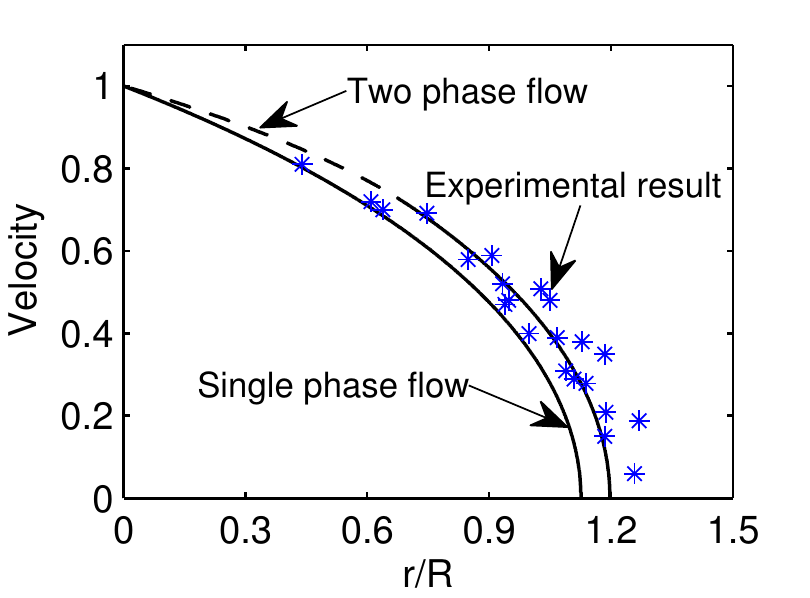}
       \hspace{0.2in}\parbox{3.4in}{\caption{Comparison of the velocity profiles of two phase and single phase blood flow model with  the experimental results of Bugliarello and Sevilla 
       \cite{bugliarello1970velocity} \label{fig:exp}}}
  \end{minipage}
  \hfill
  \begin{minipage}[b]{0.49\textwidth}
    \includegraphics[width=\textwidth,height=5.5cm]{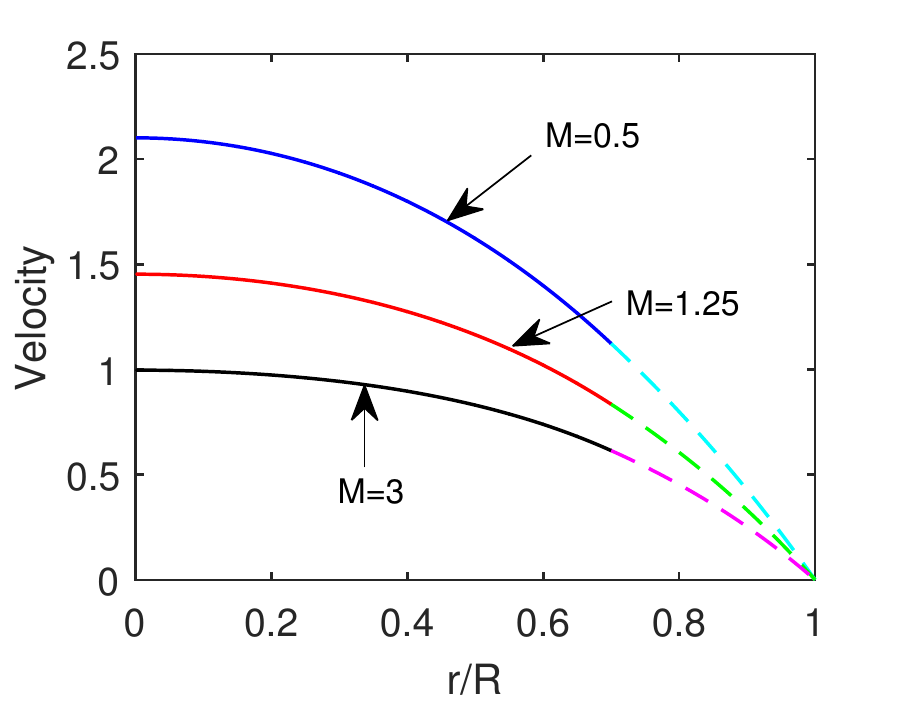}
       \hspace{0.2in}\parbox{3.4in}{\caption{Variation of velocity profile of two phase blood flow with different values of magnetic field parameter ($M$), where $\delta=0.1$, $t=1$, $R_1=0.7$ }\label{V_M}}
  \end{minipage}
\end{figure}

     \figref{fig:exp} displays the comparative studies between single phase (where we assume that plasma and red blood cells are uniformly distributed over the region and their flow dynamics are also same) and two-phase model of the blood flow with the experimental results of Bugliarello and Sevilla\cite{bugliarello1970velocity},  through their \textit{In Vitro} experimental study under the steady flow conditions measured the cell velocity distribution in a fine glass tube for $40\%$ RBC containing blood. The results of the present model are closer to those of Bugliarello and Sevilla for values $\delta=0$, $Re=5$, $Pe=1$, $M=0$ and $R_1=0.7$. Comparison of results shows that the two-phase blood flow model ($R_1=0.7$) has a good agreement with the experimental result as compared to single phase data ($R_1=1$). 
     
    The comparative data of mean squared errors (MSE) for two-phase and single-phase blood flow with the experimental data are given as

\begin{align}\label{Y_T}
\frac{1}{n}{\sum^n_{i=1}}{{(\widehat{\textbf{Y}}_{\textbf{EX}_i}-{\textbf{Y}_{\textbf{T}_i}}})^2}=0.003094882,
\end{align} 
\begin{align}\label{Y_S}
\frac{1}{n}{\sum^n_{i=1}}{{(\widehat{\textbf{Y}}_{\textbf{EX}_i}-{\textbf{Y}_{\textbf{S}_i}}})^2}=0.010313676,
\end{align}
 where $\widehat{\textbf{Y}}_{\textbf{EX}}$ represents the experimental data, $\textbf{Y}_{\textbf{T}}$ is the two-phase model data and $\textbf{Y}_{\textbf{S}}$ is single phase model data. It is noted that the two-phase model data fit with the experimental data more appropriately than the single phase model data as it has $0.3\%$ mean squared error while the single phase model data has $1\%$ mean squared error.
 
  \figref{V_M} shows the effect of the applied magnetic field on the velocity profile of the blood flow. It is observed from the figure that as the values of applied magnetic field increase from $0.5$ to $3$,  velocity profiles for both core and plasma regions decrease, respectively. This happens due to the 
  \begin{figure}[h]
  \centering
  \begin{minipage}[b]{0.49\textwidth}
    \includegraphics[width=\textwidth,height=5.5cm]{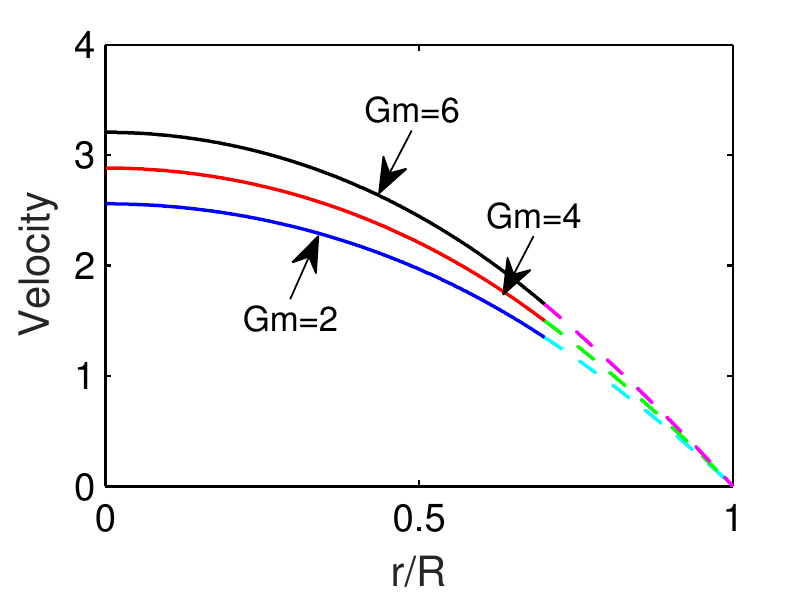}
       \hspace{0.2in}\parbox{3.4in}{\caption{Variation of velocity profile for different values of $Gm$}\label{fig:V_Gm}}
  \end{minipage}
  \hfill
  \begin{minipage}[b]{0.49\textwidth}
    \includegraphics[width=\textwidth,height=5.5cm]{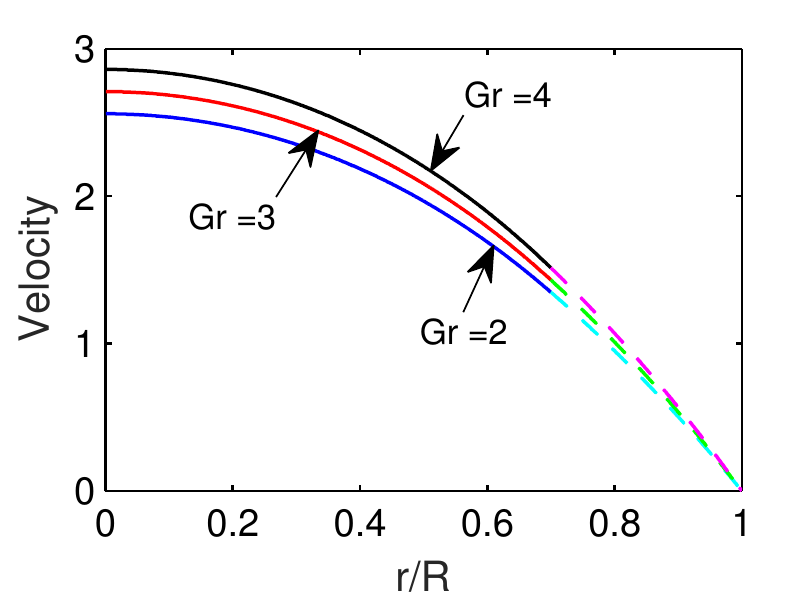}
       \hspace{0.2in}\parbox{3.4in}{\caption{Variation of velocity profile for different values of $Gr$}\label{V_Gr}}
  \end{minipage}
\end{figure}  
mature red blood cells which contain the high concentration of hemoglobin molecules in their content that basically are the oxides of iron. Therefore, when blood flows under the influence of a magnetic field, erythrocytes orient with their own disk plane parallel to the direction of the applied magnetic field. This action increases the concentration of red blood cells and causes an increase in the internal blood viscosity as discussed by Haik \textit{et al.} \cite{haik2001apparent}. Also, as boundary condition shows the continuous behavior of the flow at the interface, the reduced velocity of the red blood cells direct affects the velocity of the plasma by resulting angular velocity for the plasma fluid.  The difference between the angular velocities of plasma and red blood cell create the viscous torque which results in the decreased velocity of the plasma region as well.  This leads to a decrease in the blood velocity as Lorentz force opposes the flow of blood \cite{tzirtzilakis2005mathematical}. 
     \figref{fig:V_Gm} and \figref{V_Gr} show the variation of axial velocity profiles with solutal Grashof number and thermal Grashof number, respectively.  It is observed from the figures that as we increase the value of solutal Grashof number and thermal Grashof number, velocity profiles increase respectively for both core and plasma regions. 

\begin{figure}[h]
  \centering
  \begin{minipage}[b]{0.49\textwidth}
    \includegraphics[width=\textwidth,height=5.5cm]{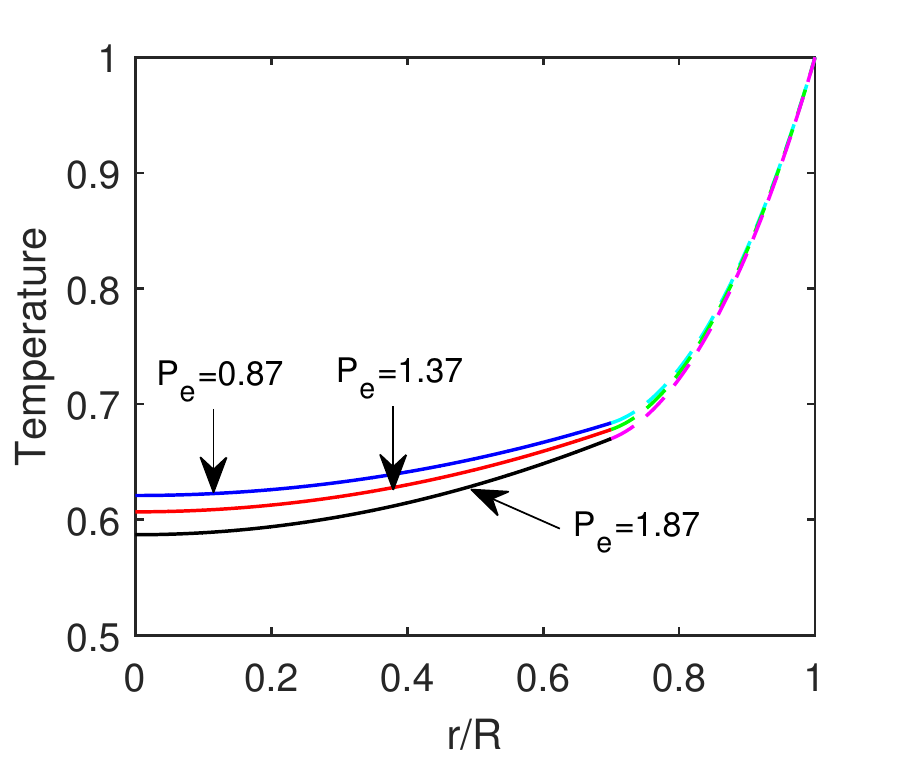}
    \setlength{\abovecaptionskip}{3ex}
       \hspace{0.2in}\parbox{3.4in}{\caption{Variation of temperature profile for different values of $Pe$}\label{TwoT_Pe}}
       
  \end{minipage}
  \hfill
  \begin{minipage}[b]{0.49\textwidth}
    \includegraphics[width=\textwidth,height=5.5cm]{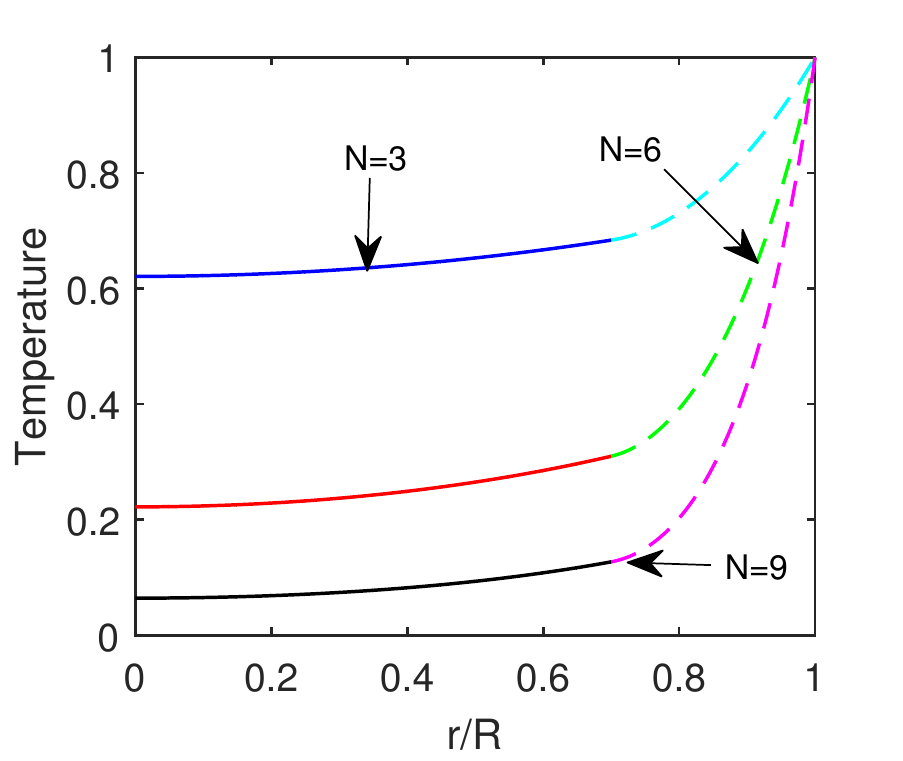}
    \setlength{\abovecaptionskip}{3ex}
       \hspace{0.2in}\parbox{3.4in}{\caption{Variation of temperature profile for different values of $N$}\label{TwoT_N}}
       
  \end{minipage}
\end{figure}
\begin{figure}[h]
  \centering
  \begin{minipage}[b]{0.49\textwidth}
   \includegraphics[width=\textwidth,height=5.5cm]{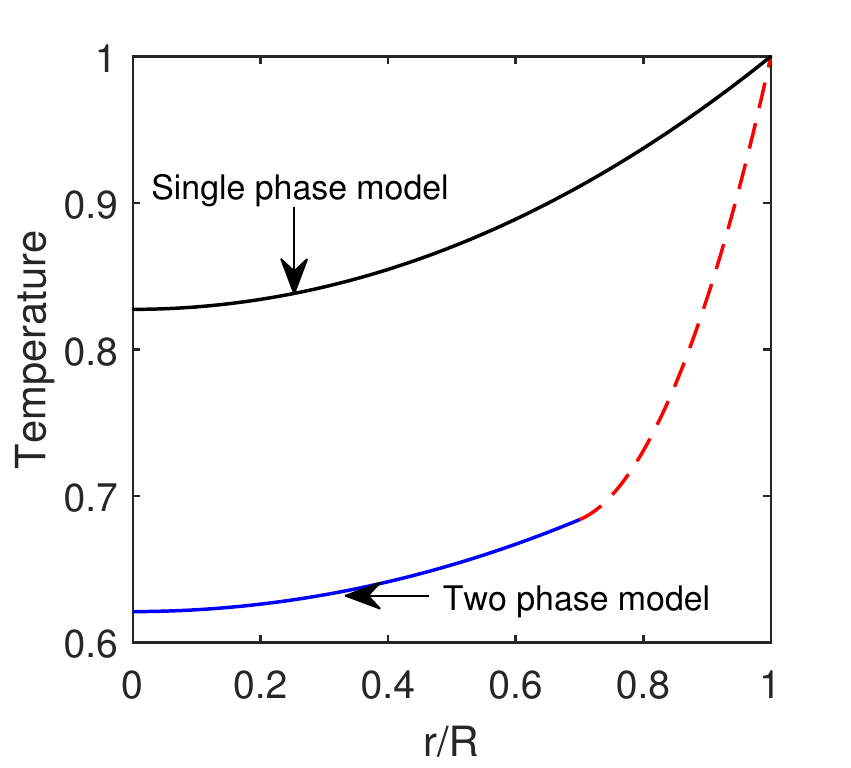}
    \setlength{\abovecaptionskip}{3ex}
       \hspace{0.2in}\parbox{3.4in}{\caption{Comparision of temperature profiles of single and two phase model of blood flow }\label{CompareT}}
  \end{minipage}
  \hfill
  \begin{minipage}[b]{0.49\textwidth}
   \includegraphics[width=\textwidth,height=5.5cm]{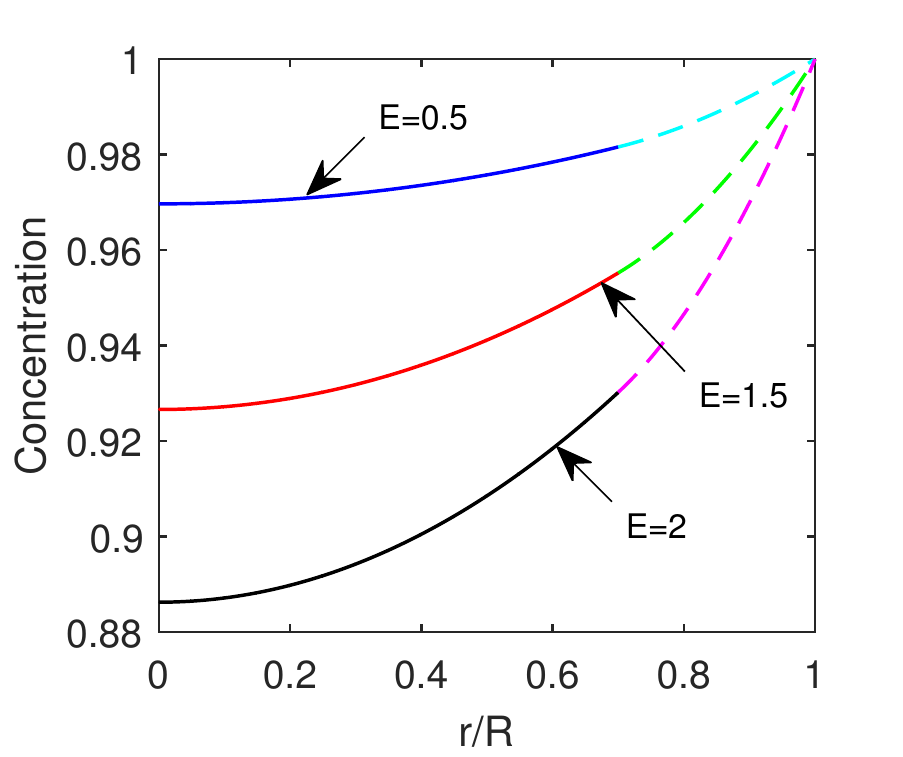}
    \setlength{\abovecaptionskip}{3ex}
       \hspace{0.2in}\parbox{3.4in}{\caption{Effect of chemical reaction parameter ($E$) on concentration profile of two phase blood flow}\label{TwoC_E}}
  
  \end{minipage}
\end{figure}
 \figref{TwoT_Pe} illustrates the variations of the temperature profile of the blood flow for different values of the Peclet number ($Pe$). It is noticed from the figure that as values of the Peclet number increase, temperature profile decreases in both core as well as plasma regions. \figref{TwoT_N} illustrates the behavior of the temperature profile of the blood flow for different values of the radiation parameter. The figure depicts that for a particular value of radiation parameter ($N$), temperature profile start increases from mid of the artery towards the interface region of the core and plasma regions and it continuously increases up to the arterial wall. Further, as we increase the values of the radiation parameter, temperature profile decreases in both core and plasma regions of the artery. The behavior of temperature profile in the presence of radiation parameter is same as discussed by  \cite{ogulu2005simulation}. Comparison of the temperature profile for single phase and two-phase model of blood flow have been displayed in \figref{CompareT}, taking parameters value from Table.\ref{table 1}. It is clear from the figure that temperature profile of the two-phase blood flow along the radial direction of the artery attains lower values than the temperature profile of the single phase model of the blood flow in which plasma and red blood cells are uniformly distributed over the artery.
\begin{figure}[h]
  \centering
  \begin{minipage}[b]{0.49\textwidth}
   \includegraphics[width=\textwidth,height=5.5cm]{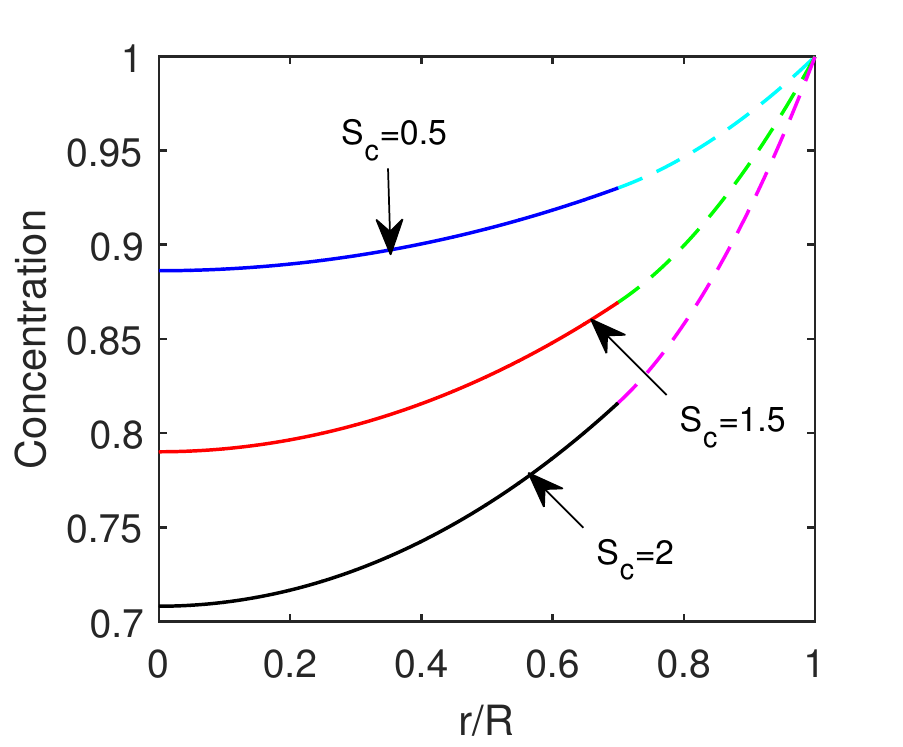}
    \setlength{\abovecaptionskip}{3ex}
       \hspace{0.2in}\parbox{3.4in}{\caption{Effects of Schmidt number ($S_c$) on concentration profile of two phase blood flow}\label{TwoC_Sc}}
  \end{minipage}
  \hfill
  \begin{minipage}[b]{0.49\textwidth}
    \includegraphics[width=\textwidth,height=5.5cm]{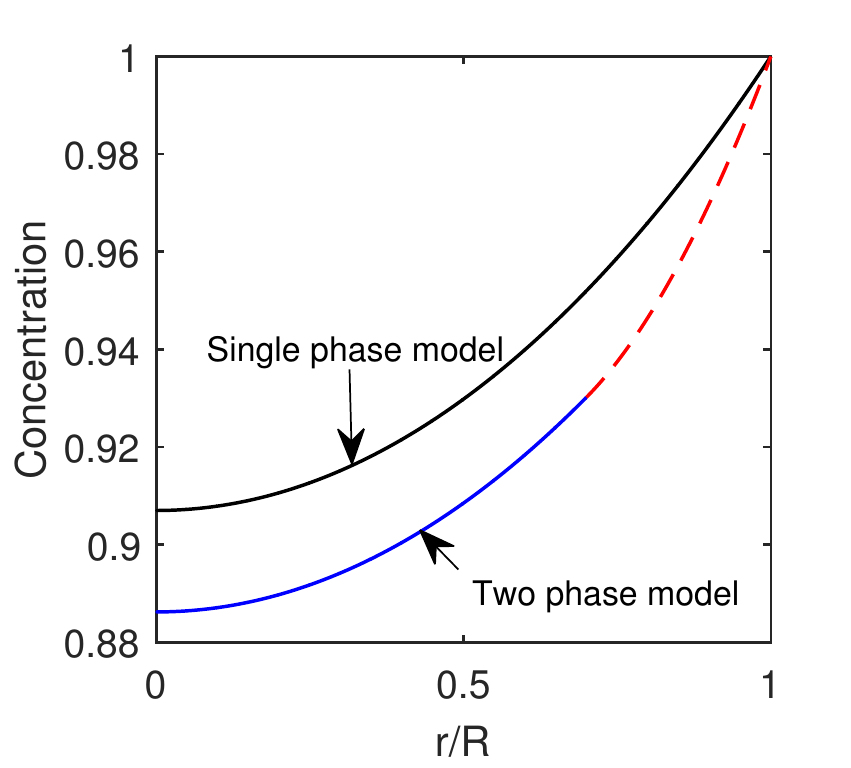}
    \setlength{\abovecaptionskip}{3ex}
       \hspace{0.2in}\parbox{3.4in}{\caption{Comparison of concentration profiles of two phase and single phase blood flow }\label{TwoCom_C}}
  \end{minipage}
\end{figure}

\figref{TwoC_E} and \figref{TwoC_Sc}  under the purview of the present computational study reveal that the concentration profiles for both core and plasma regions decrease with increasing values of the chemical reaction parameter and Schmidt number, respectively. Concentration profile increases for fixed values of chemical reaction parameter and Schmidt number as we move from the mid of the artery towards the interface region and further up to the arterial wall. Concentration profile shows this behavior because higher values of chemical reaction parameter result a fall into the molecular diffusivity which directly suppresses the species concentration. 
\figref{TwoCom_C} shows the comparison for the concentration profile of the single-phase blood flow ($R_1=1$) and for the two-phase blood flow ($R_1=0.7$). It is clear from the figure that, concentration profile along the radial direction of the artery attains higher values for single phase blood flow (in which blood components are uniformly distributed over the artery) than the two-phase model of the blood flow.
\begin{figure}[h]
  \centering
  \begin{minipage}[b]{0.49\textwidth}
    \includegraphics[width=\textwidth,height=5.5cm]{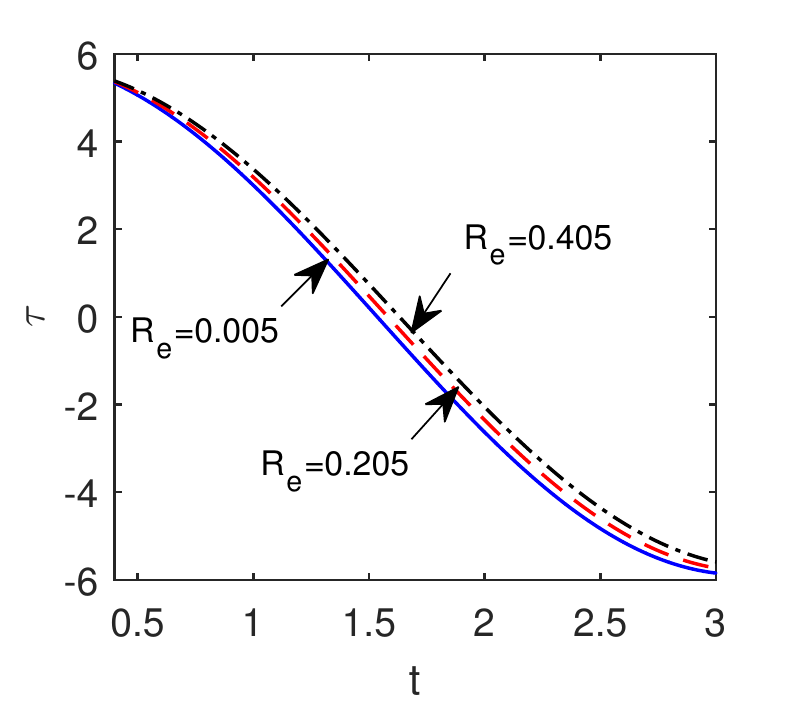}
    \setlength{\abovecaptionskip}{3ex}
       \hspace{0.2in}\parbox{3.4in}{\caption{Variation of wall shear stress with time for different values of   $Re$ }\label{Sher_Re}}
  \end{minipage}
  \hfill
  \begin{minipage}[b]{0.49\textwidth}
    \includegraphics[width=\textwidth,height=5.5cm]{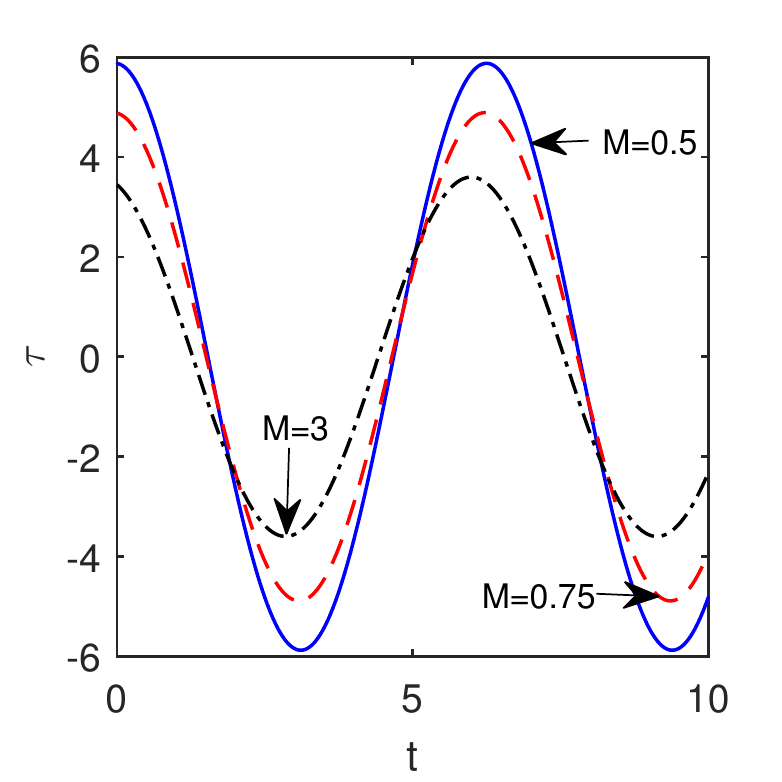}
    \setlength{\abovecaptionskip}{3ex}
       \hspace{0.2in}\parbox{3.4in}{\caption{Variation of wall shear stress with time for different values of  $M$ }\label{Sher_M}}
  \end{minipage}
\end{figure}

   In the study of stenosed arterial blood flow, wall shear stress is a major flow component to measure \cite{cavalcanti1992wall} as in the artery, vascular tissue shows histological and morphological alterations when it is physically stressed. The shear stress at the interface wall of core and the plasma region is obtained as
\begin{align}
{\tau}=\left(\frac{\partial{u_c}}{\partial{r}}\right)_{R_1}.
\end{align}
At the outer wall of the artery shear stress is determined as
\begin{align}
{\tau'}=\frac{1}{\mu_0}\left(\frac{\partial{u_p}}{\partial{r}}\right)_{R_0},
\end{align} 
where $\mu_0$ is the ratio of the viscosity in plasma and core regions, respectively.

 For unsteady blood flow \figref{Sher_Re} displays the variation of wall shear stress along with the time, for different values of the Reynolds number. It is clearly observed from the figure that as the Reynolds number increases from $0.005$ to $0.405$ along with the time cycle, shear stress at the wall of the artery also increases. Here, we consider only the plasma velocity to evaluate the wall shear stress in the stenosed artery due to the existence of the plasma layer near the arterial wall. \figref{Sher_M}  depicts the variation of wall shear stress along with time for different values of the applied magnetic field.  From the figure, it is clear that for different values of the magnetic field parameter wall shear stress shows the oscillatory behavior due to the pulsatile nature of the blood flow.
\begin{figure}[h]
  \centering
  \begin{minipage}[b]{0.49\textwidth}
    \includegraphics[width=\textwidth,height=5.5cm]{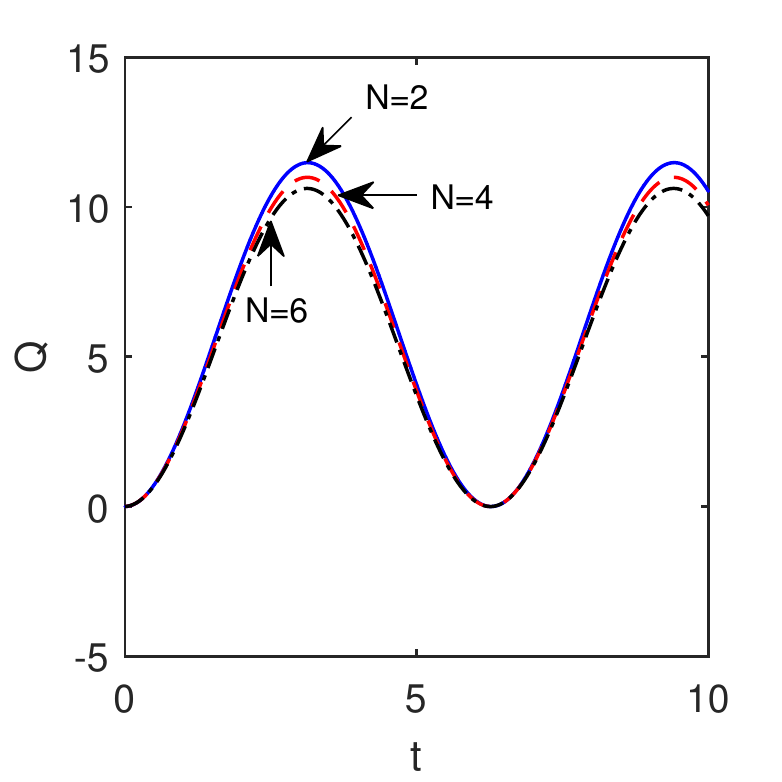}
    \setlength{\abovecaptionskip}{3ex}
       \hspace{0.2in}\parbox{3.4in}{\caption{Variation of total flow rate with time for different values of $N$}\label{Flow_N}}
  \end{minipage}
  \hfill
  \begin{minipage}[b]{0.49\textwidth}
    \includegraphics[width=\textwidth,height=5.5cm]{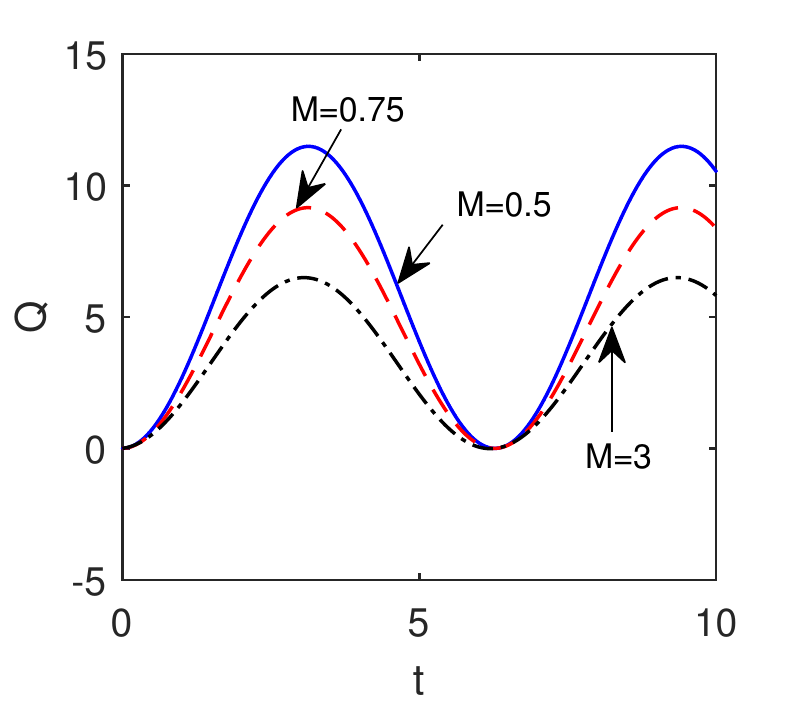}
    \setlength{\abovecaptionskip}{3ex}
       \hspace{0.2in}\parbox{3.4in}{\caption{ Variation of total flow rate with time for different values of  $M$ }\label{Flow_M}}
  \end{minipage}
\end{figure}
\begin{figure}[h]
  \centering
  \begin{minipage}[b]{0.49\textwidth}
    \includegraphics[width=\textwidth,height=5.5cm]{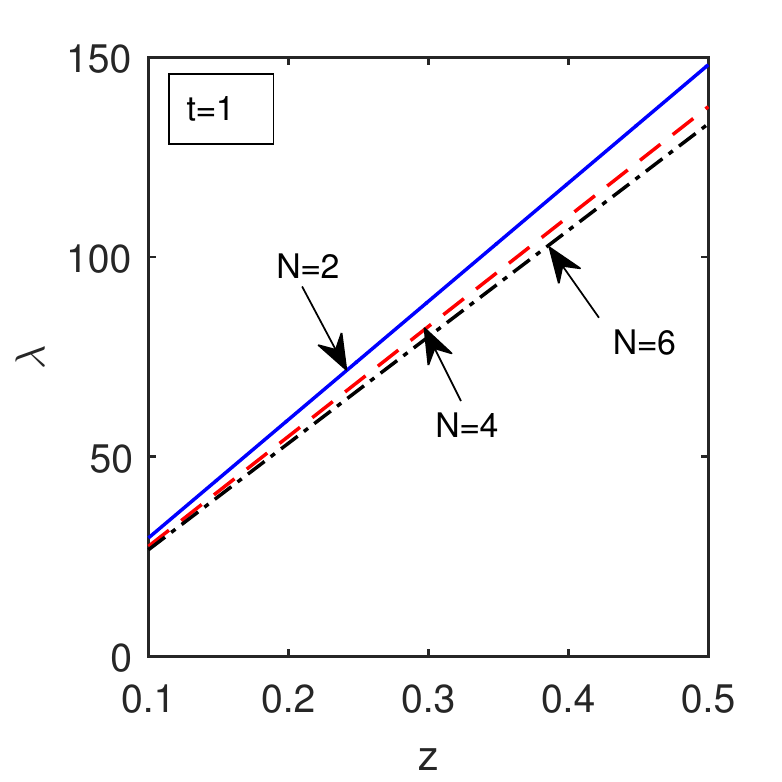}
    \setlength{\abovecaptionskip}{3ex}
       \hspace{0.2in}\parbox{3.4in}{\caption{Variation of flow impedence with axial distance for different values of $N$}\label{Lambda_N}}
  \end{minipage}
  \hfill
  \begin{minipage}[b]{0.49\textwidth}
    \includegraphics[width=\textwidth,height=5.5cm]{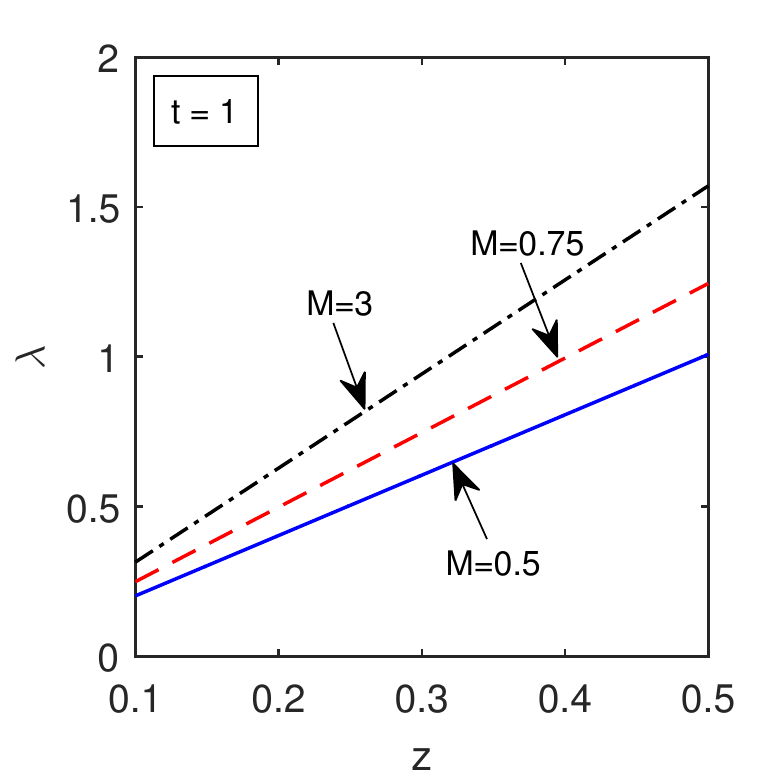}
    \setlength{\abovecaptionskip}{3ex}
       \hspace{0.2in}\parbox{3.4in}{\caption{Variation of flow impedence with axial distance for different values of  $M$}\label{Impedence_M}}
  \end{minipage}
\end{figure}

 The total volumetric flow rate of the blood flow in the artery is calculated as
\begin{align}\label{Q}
{Q}=2{\pi}{R^2}\int_{0}^{R_1}{{u_c}(r,t)}dr+2{\pi}{R^2}\int_{R_1}^{R}{{u_p}(r,t)}dr.
\end{align}
\figref{Flow_N} and \figref{Flow_M} show the flow rate profile of the blood flow along with time $t$ for different values of the radiation and magnetic parameter, respectively. These figures reveal that for increasing values of the radiation and magnetic field parameter, flow rate decreases. Flow rate displays this behavior with magnetic parameter because blood behaves as electrically conducting fluid which induces electric as well as the magnetic field when it flows under the influence of the magnetic field. Therefore, the combined effect of both the forces produces a body force known as ``Lorentz force", which has the tendency to oppose the fluid motion \cite{shit2016effect}.

 The flow impedance gives the strong correlation between the localization of stenosis and arterial wall as it is important to understand the development of arterial disease. Flow resistance in two-phase blood flow is calculated as
\begin{align}
\lambda={\int_0^z}\frac{{P_0}{e^{i\omega{t}}}}{Q}dz,
\end{align}
\figref{Lambda_N} and \figref{Impedence_M}  show the impedance profile against the axial distance for different values of the radiation and magnetic parameter, respectively. It is observed that as values of the radiation parameter increase from $2$ to $6$, the total impedance of the blood flow decreases while the reverse effect is observed for the magnetic parameter.

\begin{figure}[h]
  \centering
  \begin{minipage}[b]{0.49\textwidth}
    \includegraphics[width=\textwidth,height=5.5cm]{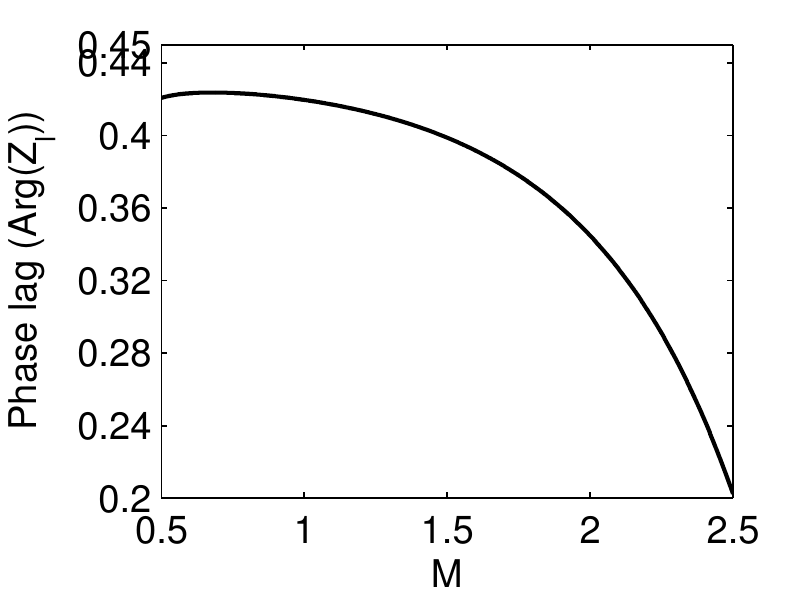}
    \setlength{\abovecaptionskip}{3ex}
       \hspace{0.2in}\parbox{3.4in}{\caption{Phase lag between pressure gradient and flow resistance with  $M$ }\label{PhaseM}}
  \end{minipage}
  \hfill
  \begin{minipage}[b]{0.49\textwidth}
    \includegraphics[width=\textwidth,height=5.5cm]{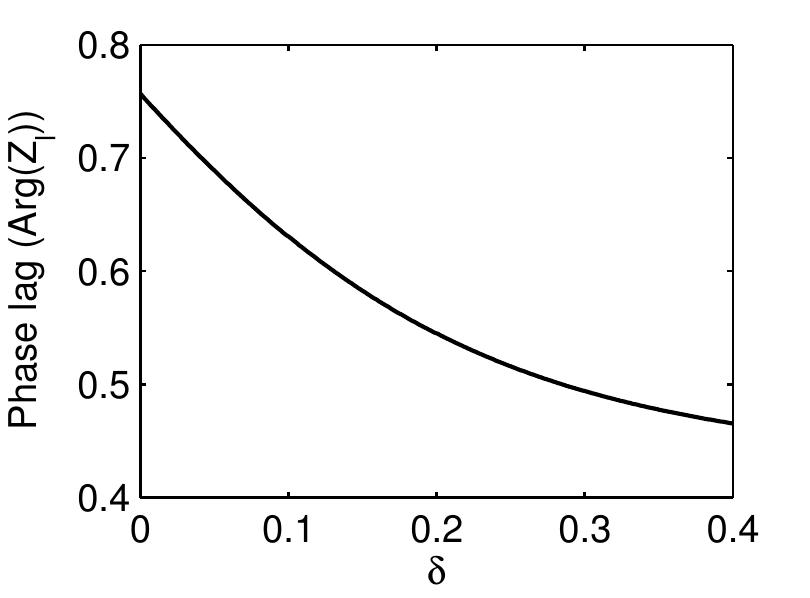}
    \setlength{\abovecaptionskip}{3ex}
       \hspace{0.2in}\parbox{3.4in}{\caption{Phase lag between pressure gradient and flow resistance with  $\delta$ }\label{PhaseD}}
  \end{minipage}
\end{figure}

 Now, to observe the phase difference between pressure gradient and flow rate for pulsatile flow,  expression of the longitudinal impedance (which relates the forces acting on blood due to the local pressure gradient with the movement of the blood) is used  as
\begin{align}
Z_l=-\frac{1}{Q}\left(\frac{\partial p}{\partial z}\right)
\end{align}
where $Q$ is the flow rate. For pulsatile flow, impedance function is very important in the analysis of wave propagation, reflection of pressure and flow pulses traveling through an arterial system. 
  \figref{PhaseM} and \figref{PhaseD} display the phase difference between the pressure gradient and flow rate with the magnetic field parameter and the variable height of the stenosis, respectively. \figref{PhaseM} shows that $Arg(Z_l)$ decreases with increase in the magnetic field parameter ($M$). For smaller values of magnetic field parameter phase difference between the pressure gradient and the flow rate is $\frac{\pi}{12}$. From the same figure, it is concluded that for higher values of the magnetic field parameter ($M>0.5$) phase difference is decreasing between pressure gradient and flow rate. \figref{PhaseD} illustrates that $Arg(Z_l)$ significantly decreases as the height of the stenosis increases. Further, the figure reveals that the pressure gradient and flow rate are in the phase difference of $\frac{\pi}{4}$ in the absence of stenosis in the arterial segment and it slowly decreases as the height of the stenosis increases.
 \begin{figure}[h]
  \centering
  \begin{minipage}[b]{0.322\textwidth}
    \includegraphics[width=\textwidth,height=5.25cm]{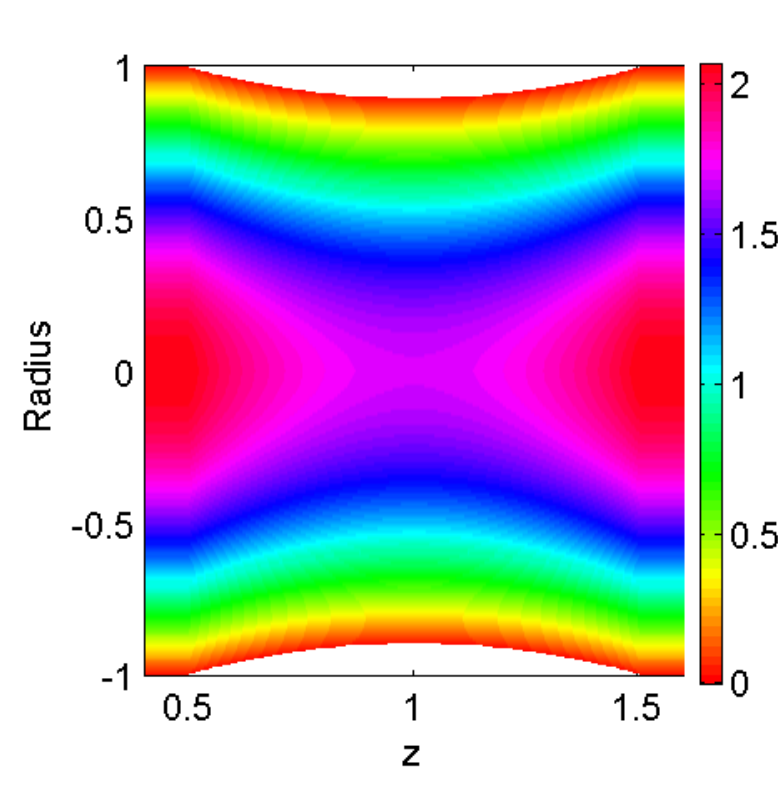}
    \setlength{\abovecaptionskip}{3ex}
       \hspace{1in}\parbox{2.5in}{\caption{Two phase blood flow analysis for $10\%$ stenosis }\label{ContourV_D1}}
  \end{minipage}
  \hfill
  \begin{minipage}[b]{0.33\textwidth}
    \includegraphics[width=\textwidth,height=5.25cm]{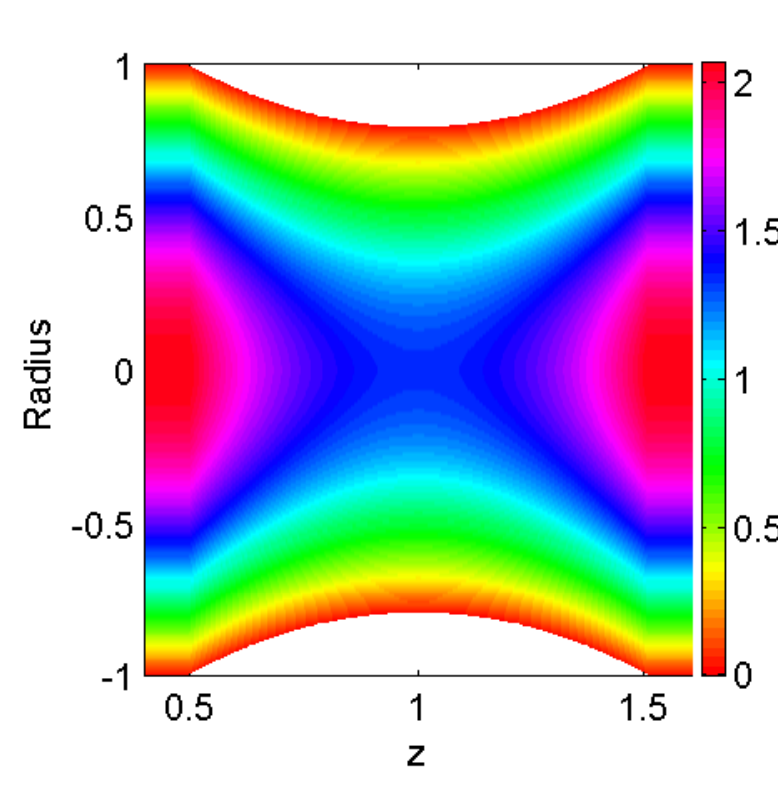}
    \setlength{\abovecaptionskip}{3ex}
       \hspace{0.05in}\parbox{2.5in}{\caption{Two phase blood flow analysis for $20\%$ stenosis }\label{ContourV_D2}}
  \end{minipage}
  \begin{minipage}[b]{0.33\textwidth}
    \includegraphics[width=0.9\textwidth,height=5.25cm]{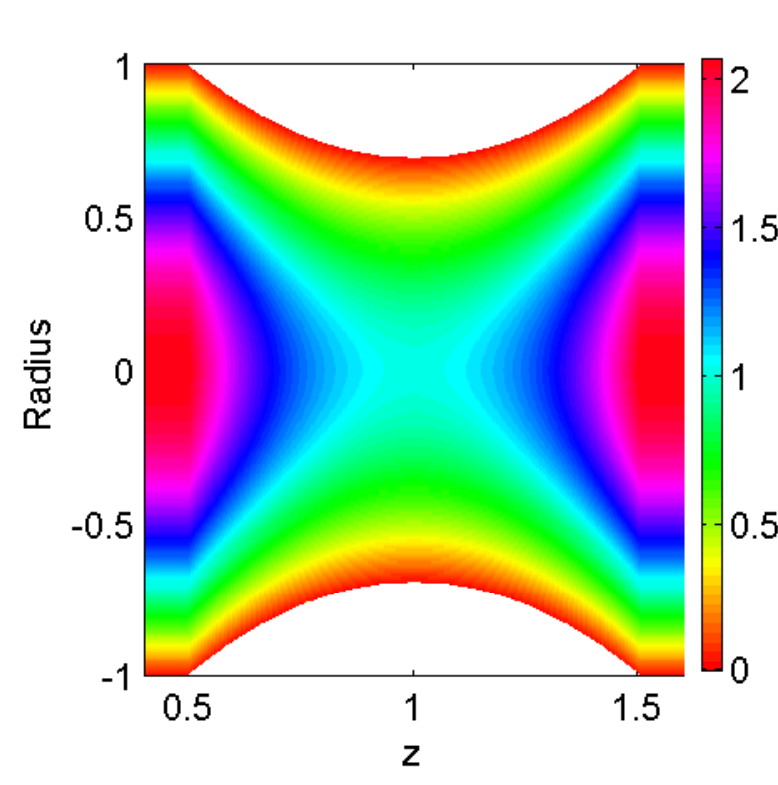}
    \setlength{\abovecaptionskip}{3ex}
       \hspace{0.05in}\parbox{2.5in}{\caption{Two phase blood flow analysis for $30\%$ stenosis}\label{ContourV_D3}}
  \end{minipage}
  \hfill
\end{figure}
  \begin{figure}[h]
  \centering
  \begin{minipage}[b]{0.322\textwidth}
    \includegraphics[width=\textwidth,height=5.25cm]{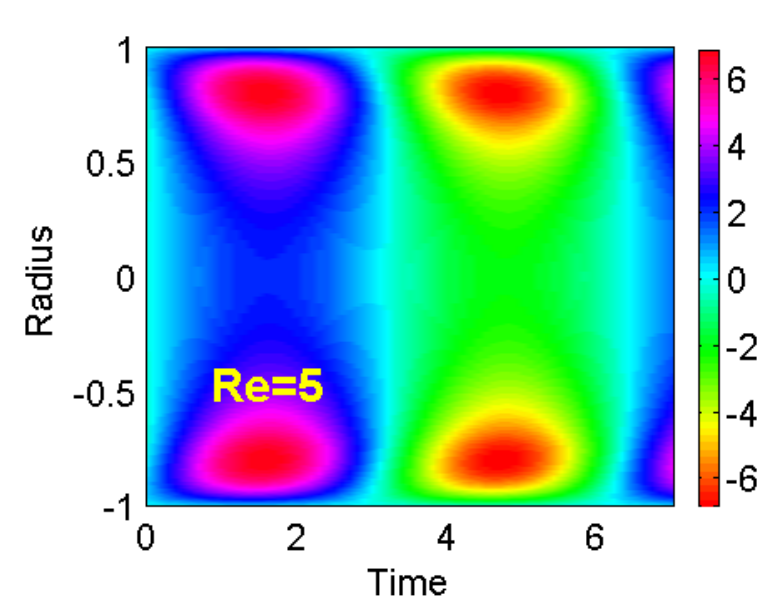}
    \setlength{\abovecaptionskip}{3ex}
       \hspace{1in}\parbox{2.5in}{\caption{Two phase blood flow analysis at ${R_e}=10$}\label{ContourV_Re1}}
  \end{minipage}
  \hfill
  \begin{minipage}[b]{0.33\textwidth}
    \includegraphics[width=\textwidth,height=5.25cm]{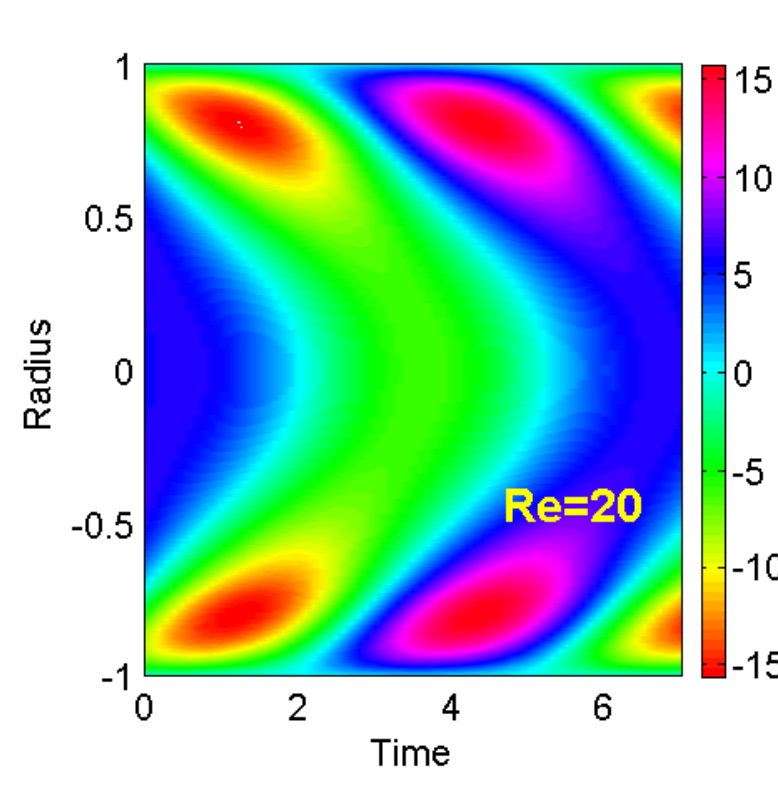}
    \setlength{\abovecaptionskip}{3ex}
       \hspace{0.05in}\parbox{2.5in}{\caption{Two phase blood flow analysis at ${R_e}=30$}\label{ContourV_Re2}}
  \end{minipage}
  \begin{minipage}[b]{0.33\textwidth}
    \includegraphics[width=0.9\textwidth,height=5.25cm]{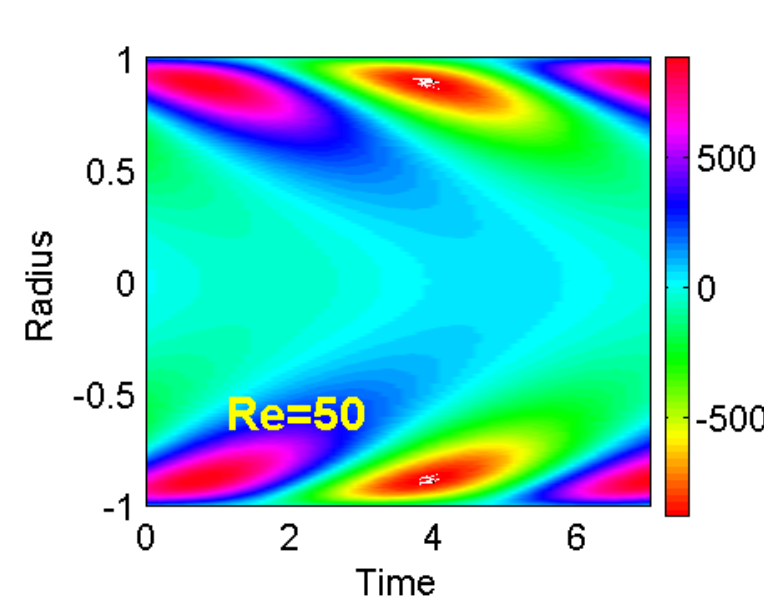}
    \setlength{\abovecaptionskip}{3ex}
       \hspace{0.05in}\parbox{2.5in}{\caption{Two phase blood flow analysis at ${R_e}=50$ }\label{ContourV_Re3}}
  \end{minipage}
  \hfill
\end{figure}

 \figref{ContourV_D1} to \figref{ContourV_Re3} emphasize on displaying the contour plots for the two-phase model of the blood flow through the constricted part of the artery under the action of an applied magnetic field. Under a range of hemodynamic flow, all the contour plots have been plotted using the data given in Table.\ref{table 1}. \figref{ContourV_D1} to \figref{ContourV_D3} show the velocity contour along the axial direction of the artery as blockage of the artery increases while considering the fixed value of the applied magnetic field as $M=1$.  These figures illustrate the flow patterns of blood in different positions of the artery, viz., entry section, onset, throat and downstream of the stenosis, outlet, where diseased part of the artery varies from $z=0.5$ to $z=1.5$ along the artery.  It can be clearly noticed from the figures that as blockage of the artery increases the velocity contours strongly gets distorted at the downstream and slowly appearing trapping bolus shift toward the arterial wall, thereby velocity decreases.  
  For unsteady flow, \figref{ContourV_Re1} to \figref{ContourV_Re3} exhibit the flow patterns of blood through the contours of the two-phase blood flow against time for different values of the Reynolds number. These figures capture flow circulation at 10\% constriction of
 the artery when Reynold number varies from $Re=5$ to $Re=50$. It is observed that the region of flow circulation is small when $Re=5$ and flow become turbulent for high values of Reynold number and it makes flow locally unstable with respect to large-scale of disturbances as discussed by Ahmed and Giddens \cite{ahmed1983velocity}.
\section*{Conclusion}
The main focus of the present study is
to investigate the combined effects of the plasma layer
thickness, heat and mass transfer on the blood flow through narrow stenosed arteries. Effects of different parameters such as magnetic field, radiation and chemical reaction on flow have been presented for core and plasma regions separately. Important findings obtained from the graphical results are listed herewith:
\begin{enumerate}
\item The velocity of the blood flow in narrow arteries decreases as values of the magnetic field parameter increase and velocity of plasma region attains lower value than the core region. The given result is very much applicable in the medical field, since during the surgical process in narrow arteries blood flow can be regulated at the desired level.  
\item Increase in the peripheral plasma layer thickness leads to decrease velocity, temperature and concentration profile of the blood flow.
\item The temperature of the blood flow reduces to an appreciable extent in both core and plasma regions, as the values of the radiation parameter increase. Hence, the present study reveals that the temperature of blood flow can be regulated in the narrow arteries by reducing or increasing the effects of the radiation parameter. The result is very much useful in radiation therapy which is used to treat cancer. 
\item The concentration profile decreases with increasing  the chemical reaction parameter. This happens due to increased molecular diffusivity which directly suppresses the concentration profile of the flow.
\item The investigation shows that Peclet and Schmidt number have reducing effects on temperature and concentration of the blood flow, respectively.
\item The high intensity of magnetic field causes reduction of the flow rate while it has an enhancing effect on flow impedance. 
\item  Both flow rate and flow impedance decrease as values of the radiation parameter increase.
\item A comparative study between the present result and the experimental result of the cell velocity distribution of $40\%$ RBC containing blood validate the present model. The comparative result shows that the present result gives the good agreement with the experimental results.
\item For pulsatile flow, the phase difference between the pressure gradient and flow rate decreases with applied magnetic field and height of the stenosis.
\end{enumerate}
\section*{Acknowledgment}
 Authors are sincerely thankful to the Department of Science and Technology, Government of India (SR/FST/MSI-090/2013(C) ) for their financial support.
\appendix\label{Appendix}
\section*{Appendix A}
\begin{appendix}
\begin{align*}
{\beta_1}=-\left(\frac{{K_0}{{N}^2}}{{\alpha_0}}+{i\frac{P_e}{\rho_0}\left(\frac{K_0}{s_0}\right)}\right), \quad \beta_2 = -\left({{N}^2}+i{P_e}\right).
\end{align*}
\begin{align*}
{\gamma_1} =-\left(i{Re}{D_0}{S_c}+\frac{E}{E_0}{D_0}{S_c}\right), \quad {\gamma_2} = -\left(i{Re}{S_c}+{E}{S_c}\right)
\end{align*}
\begin{align}
\lambda_1 = -\left(M^2+\frac{{\mu_0}{Re}}{\rho_0}i\right) \quad \lambda_2= -\left(M^2+{{Re}i}\right)
\end{align}

\begin{align*}
{U_1}=\left(\frac{{J_0}({\sqrt{\beta_2}}{R_1})}{{J_0}({\sqrt{\beta_1}}{R_1})}-\frac{{J_0}({\sqrt{\beta_2}}{R})}{{Y_0}({\sqrt{\beta_2}}{R})}{\frac{{Y_0}({\sqrt{\beta_2}}{R_1})}{{J_0}({\sqrt{\beta_1}}{R_1})}}\right), {U_2}=\frac{{Y_0}({\sqrt{\beta_2}}{R_1}))}{{J_0}({\sqrt{\beta_1}}{R_1}){Y_0(\sqrt{\beta_2}R)}}
\end{align*}
\begin{align*}
 {U_3}={\sqrt{\beta_1}}{U_1}{J_1}(\sqrt{\beta_1}{R_1})-\sqrt{\beta_2}\left({{{J_1}({\sqrt{\beta_2}}{R_1})}-{\frac{{J_0}({\sqrt{\beta_2}}{R})}{{Y_0(\sqrt{\beta_2}R)}}{{Y_1}}({\sqrt{\beta_2}}{R_1})}}\right)
\end{align*} 
\begin{align*}
{U_4}=\left(\frac{{J_0}({\sqrt{\gamma_2}}{R_1})}{{J_0}({\sqrt{\gamma_1}}{R_1})}-\frac{{J_0}({\sqrt{\gamma_2}}{R})}{{J_0}({\sqrt{\gamma_1}}{R_1})}{\frac{{Y_0}({\sqrt{\gamma_2}}{R_1})}{{Y_0}({\sqrt{\gamma_2}}{R})}}\right), {U_5}=\frac{{Y_0}({\sqrt{\gamma_2}}{R_1}))}{{J_0}({\sqrt{\gamma_1}}{R_1}){Y_0(\sqrt{\gamma_2}R)}}
\end{align*}
\begin{align*}
{U_6}={\sqrt{\gamma_1}}{U_5}{J_1}(\sqrt{\gamma_1}{R_1})-\sqrt{\gamma_2}{\left({{J_1}({\sqrt{\gamma_2}}{R_1})}-{\frac{{J_0}({\sqrt{\gamma_2}}{R})}{{Y_0(\sqrt{\gamma_2}R)}}.{{Y_1}({\sqrt{\gamma_2}}{R_1})}} \right)}
\end{align*}
\begin{align*}
D_1=-{A_1}(R_1){J_0}(\sqrt{\lambda_1}{R_1})-{B_1}(R_1){Y_0}(\sqrt{\lambda_1}{R_1})+{A_2}(R_1){J_0}(\sqrt{\lambda_2}{R_1})+{B_2}(R_1){Y_0}(\sqrt{\lambda_2}{R_1})
\end{align*}
\begin{align*}
D_2 = -{A_2}(R){J_0}(\sqrt{\lambda_2}{R})-{B_2}(R){Y_0}(\sqrt{\lambda_2}{R})
\end{align*}
\begin{align*}
D_3= -\frac{\partial{A_1(R_1)}}{\partial{r}} {J_0}(\sqrt{\lambda_1}{R_1})+{A_1}(R_1){\sqrt{\lambda_1}}{J_1}(\sqrt{\lambda_1}{R_1}) -\frac{\partial{B_1(R_1)}}{\partial{r}} {Y_0}(\sqrt{\lambda_1}{R_1})
\end{align*}
\begin{align*}
+{B_1}(R_1){\sqrt{\lambda_1}}{Y_1}(\sqrt{\lambda_1}{R_1})+\frac{\partial{A_2(R_1)}}{\partial{r}} {J_0}(\sqrt{\lambda_2}{R_1})+\frac{\partial{B_2(R_1)}}{\partial{r}} {Y_0}(\sqrt{\lambda_2}{R_1})
\end{align*}
\begin{align*}
& -{A_2}(R_1){\sqrt{\lambda_2}}{J_1}(\sqrt{\lambda_2}{R_1})-{B_2}(R_1){\sqrt{\lambda_2}}{Y_1}(\sqrt{\lambda_2}{R_1})
\end{align*}
\end{appendix}
\bibliographystyle{unsrt}
\bibliography{BibData}

\begin{thebibliography}{10}

\bibitem{medvedev2011two}
AE~Medvedev and VM~Fomin.
\newblock Two-phase blood-flow model in large and small vessels.
\newblock In {\em Doklady Physics}, volume~56, pages 610--613. Springer, 2011.

\bibitem{barbee1971fahraeus}
James~H Barbee and Giles~R Cokelet.
\newblock The fahraeus effect.
\newblock {\em Microvascular research}, 3(1):6--16, 1971.

\bibitem{faahraeus1931viscosity}
Robin F{\aa}hr{\ae}us and Torsten Lindqvist.
\newblock The viscosity of the blood in narrow capillary tubes.
\newblock {\em American Journal of Physiology--Legacy Content}, 96(3):562--568,
  1931.

\bibitem{verma1analytical}
SR~Verma.
\newblock Analytical study of a two-phase model for steady flow of blood in a
  circular tube.
\newblock {\em International Journal of engineering Research and Applications},
  1(4):1--10, 2014.

\bibitem{cokelet1991decreased}
Giles~R Cokelet and Harry~L Goldsmith.
\newblock Decreased hydrodynamic resistance in the two-phase flow of blood
  through small vertical tubes at low flow rates.
\newblock {\em Circulation research}, 68(1):1--17, 1991.

\bibitem{sharan2001two}
Maithili Sharan and Aleksander~S Popel.
\newblock A two-phase model for flow of blood in narrow tubes with increased
  effective viscosity near the wall.
\newblock {\em Biorheology}, 38(5, 6):415--428, 2001.

\bibitem{ku1997blood}
David~N Ku.
\newblock Blood flow in arteries.
\newblock {\em Annual Review of Fluid Mechanics}, 29(1):399--434, 1997.

\bibitem{alexopoulos2014visceral}
Nikolaos Alexopoulos, Demosthenes Katritsis, and Paolo Raggi.
\newblock Visceral adipose tissue as a source of inflammation and promoter of
  atherosclerosis.
\newblock {\em Atherosclerosis}, 233(1):104--112, 2014.

\bibitem{ponalagusamy2016two}
R~Ponalagusamy.
\newblock Two-fluid model for blood flow through a tapered arterial stenosis:
  Effect of non-zero couple stress boundary condition at the interface.
\newblock {\em International Journal of Applied and Computational Mathematics},
  38(2):1--18, 2016.

\bibitem{sankar2011two}
DS~Sankar.
\newblock Two-phase non-linear model for blood flow in asymmetric and
  axisymmetric stenosed arteries.
\newblock {\em International Journal of Non-Linear Mechanics}, 46(1):296--305,
  2011.

\bibitem{haik2001apparent}
Yousef Haik, Vinay Pai, and Ching-Jen Chen.
\newblock Apparent viscosity of human blood in a high static magnetic field.
\newblock {\em Journal of Magnetism and Magnetic Materials}, 225(1-2):180--186,
  2001.

\bibitem{yadav2008experimental}
RP~Yadav, S~Harminder, and S~Bhoopal.
\newblock Experimental studies on blood flow in stenosis arteries in presence
  of magnetic field.
\newblock {\em Ultra Sci}, 20(3):499--504, 2008.

\bibitem{shit2014mathematical}
GC~Shit, M~Roy, and A~Sinha.
\newblock Mathematical modelling of blood flow through a tapered overlapping
  stenosed artery with variable viscosity.
\newblock {\em Applied Bionics and Biomechanics}, 11(4):185--195, 2014.

\bibitem{mekheimer2008effect}
Kh~S Mekheimer.
\newblock Effect of the induced magnetic field on peristaltic flow of a couple
  stress fluid.
\newblock {\em Physics Letters A}, 372(23):4271--4278, 2008.

\bibitem{mekheimer2008influence}
Kh~S Mekheimer and MA~El~Kot.
\newblock Influence of magnetic field and hall currents on blood flow through a
  stenotic artery.
\newblock {\em Applied Mathematics and Mechanics}, 29(8):1093, 2008.

\bibitem{cavalcanti1995hemodynamics}
Silvio Cavalcanti.
\newblock Hemodynamics of an artery with mild stenosis.
\newblock {\em Journal of Biomechanics}, 28(4):387--399, 1995.

\bibitem{ponalagusamy2015influence}
R~Ponalagusamy and R~Tamil Selvi.
\newblock Influence of magnetic field and heat transfer on two-phase fluid
  model for oscillatory blood flow in an arterial stenosis.
\newblock {\em Meccanica}, 50(4):927--943, 2015.

\bibitem{ponalagusamy2016numerical}
R~Ponalagusamy and S~Priyadharshini.
\newblock Numerical investigation on two-fluid model (micropolar-newtonian) for
  pulsatile flow of blood in a tapered arterial stenosis with radially variable
  magnetic field and core fluid viscosity.
\newblock {\em Computational and Applied Mathematics}, pages 1--25, 2016.

\bibitem{mirza2016transient}
IA~Mirza, M~Abdulhameed, D~Vieru, and S~Shafie.
\newblock Transient electro-magneto-hydrodynamic two-phase blood flow and
  thermal transport through a capillary vessel.
\newblock {\em Computer Methods and Programs in Biomedicine},
  137(2016):149--166, 2016.

\bibitem{sapareto1984thermal}
Stephen~A Sapareto and William~C Dewey.
\newblock Thermal dose determination in cancer therapy.
\newblock {\em International Journal of Radiation Oncology* Biology* Physics},
  10(6):787--800, 1984.

\bibitem{damianou1993focal}
C~Damianou and K~Hynynen.
\newblock Focal spacing and near-field heating during pulsed high temperature
  ultrasound therapy.
\newblock {\em Ultrasound in medicine \& biology}, 19(9):777--787, 1993.

\bibitem{szasz2007hyperthermia}
A~Szasz et~al.
\newblock Hyperthermia, a modality in the wings.
\newblock {\em Journal of Cancer Research and Therapeutics}, 3(1):56, 2007.

\bibitem{sharma2013heat}
BK~Sharma, A~Mishra, and S~Gupta.
\newblock Heat and mass transfer in magneto-biofluid flow through a non-darcian
  porous medium with joule effect.
\newblock {\em Journal of Engineering Physics and Thermophysics},
  86(4):766--774, 2013.

\bibitem{sharma2015mathematical}
BK~Sharma, M~Sharma, RK~Gaur, and A~Mishra.
\newblock Mathematical modeling of magneto pulsatile blood flow through a
  porous medium with a heat source.
\newblock {\em International Journal of Applied Mechanics and Engineering},
  20(2):385--396, 2015.

\bibitem{sinha2015electromagnetohydrodynamic}
A~Sinha and GC~Shit.
\newblock Electromagnetohydrodynamic flow of blood and heat transfer in a
  capillary with thermal radiation.
\newblock {\em Journal of Magnetism and Magnetic Materials}, 378:143--151,
  2015.

\bibitem{tripathi2018influence}
Bhavya Tripathi and Bhupendra~Kumar Sharma.
\newblock Influence of heat and mass transfer on mhd two-phase blood flow with
  radiation.
\newblock In {\em AIP Conference Proceedings}, volume 1975, page 030009. AIP
  Publishing, 2018.

\bibitem{hernandez1983alligator}
Thomas Hernandez et~al.
\newblock Alligator metabolism studies on chemical reactions in vivo.
\newblock {\em Comparative biochemistry and physiology Part B: Comparative
  biochemistry}, 74(1):1--175, 1983.

\bibitem{xu2009study}
Zhiliang Xu, Nan Chen, Shawn~C Shadden, Jerrold~E Marsden, Malgorzata~M
  Kamocka, Elliot~D Rosen, and Mark Alber.
\newblock Study of blood flow impact on growth of thrombi using a multiscale
  model.
\newblock {\em Soft Matter}, 5(4):769--779, 2009.

\bibitem{sharma2017effect}
Madhu Sharma and RK~Gaur.
\newblock Effect of variable viscosity on chemically reacting magneto-blood
  flow with heat and mass transfer.
\newblock {\em Global Journal of Pure and Applied Mathematics}, 13(3):2017,
  2017.

\bibitem{mekheimer2008micropolar}
Kh~S Mekheimer and MA~El~Kot.
\newblock The micropolar fluid model for blood flow through a tapered artery
  with a stenosis.
\newblock {\em Acta Mechanica Sinica}, 24(6):637--644, 2008.

\bibitem{sankar2007two}
DS~Sankar and Usik Lee.
\newblock Two-phase non-linear model for the flow through stenosed blood
  vessels.
\newblock {\em Journal of Mechanical Science and Technology}, 21(4):678--689,
  2007.

\bibitem{nadeem2012influence}
S~Nadeem, Noreen~Sher Akbar, T~Hayat, and Awatif~A Hendi.
\newblock Influence of heat and mass transfer on newtonian biomagnetic fluid of
  blood flow through a tapered porous arteries with a stenosis.
\newblock {\em Transport in porous media}, 91(1):81--100, 2012.

\bibitem{cogley1968differential}
AC~Cogley, SE~Gilles, and WG~Vincenti.
\newblock Differential approximation for radiative transfer in a nongrey gas
  near equilibrium.
\newblock {\em AIAA Journal}, 6(3):551--553, 1968.

\bibitem{sharma2007radiation}
Bhupendra~Kumar Sharma, Mamta Agarwal, and RC~Chaudhary.
\newblock Radiation effect on temperature distribution in three-dimensional
  couette flow with suction or injection.
\newblock {\em Applied Mathematics and Mechanics}, 28(3):309--316, 2007.

\bibitem{ogulu2005simulation}
A~Ogulu and TM~Abbey.
\newblock Simulation of heat transfer on an oscillatory blood flow in an
  indented porous artery.
\newblock {\em International communications in heat and mass transfer},
  32(7):983--989, 2005.

\bibitem{san2012dynamics}
Omer San and Anne~E Staples.
\newblock Dynamics of pulsatile flows through elastic microtubes.
\newblock {\em International Journal of Applied Mechanics}, 4(1):1250006, 2012.

\bibitem{trivedi2004multi}
RA~Trivedi, J~U-King-Im, MJ~Graves, J~Horsley, M~Goddard, PJ~Kirkpatrick, and
  JH~Gillard.
\newblock Multi-sequence in vivo mri can quantify fibrous cap and lipid core
  components in human carotid atherosclerotic plaques.
\newblock {\em European journal of vascular and endovascular surgery},
  28(2):207--213, 2004.

\bibitem{swartz2009intracranial}
RH~Swartz, SS~Bhuta, RI~Farb, R~Agid, RA~Willinsky, J~Butany, BA~Wasserman,
  DM~Johnstone, FL~Silver, DJ~Mikulis, et~al.
\newblock Intracranial arterial wall imaging using high-resolution 3-tesla
  contrast-enhanced mri.
\newblock {\em Neurology}, 72(7):627--634, 2009.

\bibitem{zaman2016heat}
A~Zaman, N~Ali, O~Anwar B{\'e}g, and M~Sajid.
\newblock Heat and mass transfer to blood flowing through a tapered overlapping
  stenosed artery.
\newblock {\em International Journal of Heat and Mass Transfer}, 95:1084--1095,
  2016.

\bibitem{misra2016mhd}
JC~Misra and SD~Adhikary.
\newblock Mhd oscillatory channel flow, heat and mass transfer in a
  physiological fluid in presence of chemical reaction.
\newblock {\em Alexandria Engineering Journal}, 55(1):287--297, 2016.

\bibitem{sharan1997two}
Maithili Sharan, Balbir Singh, and Pawan Kumar.
\newblock A two-layer model for studying the effect of plasma layer on the
  delivery of oxygen to tissue using a finite element method.
\newblock {\em Applied Mathematical Modelling}, 21(7):419--426, 1997.

\bibitem{fujiwara2009red}
Hiroki Fujiwara, Takuji Ishikawa, R~Lima, Noriaki Matsuki, Yohsuke Imai,
  H~Kaji, M~Nishizawa, and Takami Yamaguchi.
\newblock Red blood cell motions in high-hematocrit blood flowing through a
  stenosed microchannel.
\newblock {\em Journal of Biomechanics}, 42(7):838--843, 2009.

\bibitem{bugliarello1970velocity}
George Bugliarello and Juan Sevilla.
\newblock Velocity distribution and other characteristics of steady and
  pulsatile blood flow in fine glass tubes.
\newblock {\em Biorheology}, 7(2):85--107, 1970.

\bibitem{tzirtzilakis2005mathematical}
EE~Tzirtzilakis.
\newblock A mathematical model for blood flow in magnetic field.
\newblock {\em Physics of fluids}, 17(7):077103, 2005.

\bibitem{cavalcanti1992wall}
Silvio Cavalcanti.
\newblock Wall shear stress in stenotic artery.
\newblock In {\em Engineering in Medicine and Biology Society, 1992 14th Annual
  International Conference of the IEEE}, volume~2, pages 421--422. IEEE, 1992.

\bibitem{shit2016effect}
GC~Shit and M~Roy.
\newblock Effect of induced magnetic field on blood flow through a constricted
  channel: An analytical approach.
\newblock {\em Journal of Mechanics in Medicine and Biology}, 16(3):1650030,
  2016.

\bibitem{ahmed1983velocity}
Saad~A Ahmed and Don~P Giddens.
\newblock Velocity measurements in steady flow through axisymmetric stenoses at
  moderate reynolds numbers.
\newblock {\em Journal of biomechanics}, 16(7):505509--507516, 1983.

\end{thebibliography}
\end{document}